\documentclass[aps,prl,twocolumn,superscriptaddress]{revtex4-2}

\usepackage{graphicx}
\usepackage{dcolumn}
\usepackage{bm}
\usepackage{subfigure}
\usepackage{mathrsfs}
\usepackage{multirow}
\usepackage{amsmath}

\usepackage[colorlinks,
linkcolor={red!60!black!},
anchorcolor={green!80!black!},
urlcolor={blue!80!black!},
citecolor={blue!50!black!}]{hyperref}
\usepackage[table]{xcolor}

\begin{document}

\title{Ab initio structure factors for spin-dependent dark matter direct detection}

\author{B. S. Hu} 
\email{bhu@triumf.ca}
\affiliation{TRIUMF, 4004 Wesbrook Mall, Vancouver, BC V6T 2A3, Canada}
\author{J. Padua-Arg\"uelles} 
\email{jpaduaarguelles@perimeterinstitute.ca}
\affiliation{TRIUMF, 4004 Wesbrook Mall, Vancouver, BC V6T 2A3, Canada}
\affiliation{Perimeter Institute, 31 Caroline Street North, Waterloo, ON, N2L 2Y5, Canada
}
\author{S. Leutheusser}
\email{sawl@mit.edu}
\affiliation{TRIUMF, 4004 Wesbrook Mall, Vancouver, BC V6T 2A3, Canada}
\affiliation{Center for Theoretical Physics, Massachusetts Institute of Technology, Cambridge, MA 02139, USA}
\author{T. Miyagi} 
\email{tmiyagi@triumf.ca}
\affiliation{TRIUMF, 4004 Wesbrook Mall, Vancouver, BC V6T 2A3, Canada}
\author{S. R. Stroberg} 
\email{sstroberg@anl.gov}
\affiliation{Department of Physics, University of Washington, Seattle, WA 98195, USA}
\altaffiliation{Present address: Physics Division, Argonne National Laboratory, Lemont, Illinois 60439, USA}
\author{J. D. Holt}
\email{jholt@triumf.ca}
\affiliation{TRIUMF, 4004 Wesbrook Mall, Vancouver, BC V6T 2A3, Canada}
\affiliation{Department of Physics, McGill University, Montr\'eal, QC H3A 2T8, Canada}

\date{\today}

\begin{abstract}

We present converged ab initio calculations of structure factors for elastic spin-dependent WIMP scattering off all nuclei used in dark matter direct-detection searches: $^{19}$F, $^{23}$Na, $^{27}$Al, $^{29}$Si, $^{73}$Ge, $^{127}$I, and $^{129,131}$Xe. 
From a set of established two- and three-nucleon interactions derived within chiral effective field theory, we construct consistent WIMP-nucleon currents at the one-body level, including effects from axial-vector two-body currents. 
We then apply the in-medium similarity renormalization group to construct effective valence-space Hamiltonians and consistently transformed operators of nuclear responses. 
Combining the recent advances of natural orbitals with three-nucleon forces expressed in large spaces, we obtain basis-space converged structure factors even in heavy nuclei. 
Generally results are consistent with previous calculations but large uncertainties in $^{127}$I highlight the need for further study.
 
\end{abstract}

\maketitle

The nature of dark matter (DM) is perhaps one of the most important unanswered questions in physics \cite{RevModPhys.90.045002,doi:10.1146/annurev-astro-082708-101659,Undagoitia_2015,Liu-NP13-212}.
Weakly interacting massive particles (WIMPs) \cite{doi:10.1142/S0218301392000023,JUNGMAN1996195,Roszkowski_2018,Schumann_2019} remain among the most likely DM candidates, favored by many theories beyond the Standard Model of particle physics.
A large number of direct-detection experiments worldwide aim to gain some insight into their properties through observation of the nuclear recoil from the scattering of galactic WIMPs off particular target nuclei. 
In order to meaningfully interpret the results of such searches, accurate theoretical WIMP-nucleus cross sections are needed to determine the rate of potential DM scattering events. 
Such calculations are tremendously complicated and require solid theoretical underpinnings of both particle and nuclear physics: first to determine how WIMPs interact with nucleons in nuclei, then to fold in the relevant nuclear structure for detector nuclei, which span the medium- and heavy-mass regions.

Chiral effective field theory (EFT) provides a systematic expansion and consistent treatment of both nuclear forces and one- and two-body currents (2BCs) of external probes coupling to nucleons, such as WIMP-nucleus scattering~\cite{HOFERICHTER2015410}. 
Indeed 2BCs have a significant impact on both electroweak transitions in nuclei \cite{PhysRevLett.103.102502,*PhysRevLett.122.029901,PhysRevLett.106.202502,PhysRevLett.107.062501,PhysRevC.87.035503,Bacca_2014,PhysRevLett.113.262504,PhysRevLett.117.082501,Gysbers2019} and DM scattering~\cite{PhysRevD.86.103511,PhysRevD.88.083516,*PhysRevD.89.029901,PhysRevD.99.055031,PhysRevC.99.025501,PhysRevLett.122.071301}.
Nuclear physics is then encoded in the so-called structure factors \cite{doi:10.1142/S0218301392000023}. 
To date, phenomenological large-scale shell model (LSSM) approaches, including effects from 2BCs, have provided the most prominent calculations of structure factors for detector nuclei~\cite{PhysRevD.86.103511,PhysRevD.88.083516,*PhysRevD.89.029901,Hofe16SD,PhysRevD.102.074018,PhysRevD.99.055031,Hofe17DM}. 
In the LSSM, however, operators must typically be adjusted to further account for neglected many-body physics outside the valence space and improve agreement with data~\cite{BROWN2001517,RevModPhys.77.427}, but currently no WIMP-nucleus scattering data exist to compare.

Over the past decade, ab initio nuclear theory has made rapid progress in the range of masses and physics that can be treated~\cite{10.3389/fphy.2020.00379} and has  already been applied to DM scattering in light nuclei~\cite{PhysRevD.89.074505,PhysRevLett.120.152002,DAVOUDI20211,PhysRevD.95.103011,PhysRevC.96.035805,PhysRevC.99.025501}. 
All detector nuclei, however, reside in the medium- or heavy-mass region, where significant computational hurdles have so far hindered converged calculations.
Recently a novel storage scheme for three-nucleon (3N) force matrix elements was developed, which allows converged ground- and excited-state energies to the $^{208}$Pb region~\cite{takayuki,Hu21208Pb},  as well as neutrinoless double-beta decay matrix elements in $^{136}$Xe~\cite{Bell21Ge,Belley_inprep}. 
Furthermore, the natural-orbital (NAT) basis \cite{PhysRevC.99.034321,PhysRevC.102.051303,PhysRevC.103.014321} provides a complementary framework for improving basis-space convergence.
In this Letter we exploit these combined advantages in the valence-space formulation of the in-medium similarity renormalization group (VS-IMSRG) \cite{PhysRevLett.106.222502,PhysRevLett.113.142501,HERGERT2016165,PhysRevLett.118.032502,PhysRevC.102.034320,Stro19ARNPS} to obtain basis-space converged  spin-dependent structure factors for all relevant detector nuclei: $^{19}$F, $^{23}$Na, $^{27}$Al, $^{29}$Si, $^{73}$Ge, $^{127}$I, and $^{129,131}$Xe. 
While we find our results generally agree well with previous LSSM calculations, providing a consistent picture from independent theoretical approaches, large uncertainties in $^{127}$I warrant further study.
This framework can also be applied straightforwardly to neutrino and spin-independent DM scattering.

The effective Lagrangian for the axial-vector interaction of a WIMP $\chi$ with a Standard Model field is given by~\cite{doi:10.1142/S0218301392000023,JUNGMAN1996195,PhysRevD.86.103511,PhysRevD.88.083516,*PhysRevD.89.029901}
\begin{equation}
\mathcal{L}_{\chi}=-\frac{G_{\rm F}}{\sqrt{2}} \bar{\chi} \gamma^{\mu} \gamma_{5} \chi \cdot \sum_{q=u,d,s}C_{q} \bar{q} \gamma_{\mu} \gamma_{5} q,
\end{equation}
where the sum runs over the light quark fields $q$, $G_{\rm F}$ is the Fermi coupling constant, and  $C_{q}$ are WIMP-quark coupling constants.
Here the WIMP spin is assumed to be 1/2, and we neglect the pseudoscalar interaction, which is suppressed in the non-relativistic limit \cite{doi:10.1142/S0218301392000023}.
The differential cross section for spin-dependent WIMP elastic scattering off a nucleus in a ground state with total angular momentum $J$ is given by~\cite{doi:10.1142/S0218301392000023}
\begin{equation}
\frac{d \sigma}{d q^{2}}=\frac{8 G_{\mathrm{F}}^{2}}{(2 J+1) v^{2}} S_{A}(\mathbf{q}^{2}),
\end{equation}
where $\mathbf{q}$ ($q \equiv|\mathbf{q}|$) denotes the momentum transfer from nucleus to WIMP, $v$ indicates the WIMP velocity, and $S_{A}(\mathbf{q}^2)$ is the axial-vector structure factor, obtained from detailed nuclear theory calculations.

Combining one- and two-body currents to order $Q^3$ in chiral EFT, $S_{A}(\mathbf{q}^2)$ can be expressed in terms of the transverse and longitudinal nuclear response functions $\mathcal{F}_{\pm}^{\Sigma_{L}^{\prime}}\left(q\right)$ and $\mathcal{F}_{\pm}^{\Sigma_{L}^{\prime\prime}}\left(q\right)$~\cite{PhysRevD.102.074018}
\begin{equation}
\begin{aligned}
S_{A}&= \sum_{L}\left[a_{0}\mathcal{F}_{+}^{\Sigma_{L}^{\prime}}\left(q\right)+a_{1}\left(1+\delta^{\prime}\left(\mathbf{q}^{2}\right)\right) \mathcal{F}_{-}^{\Sigma_{L}^{\prime}}\left(q\right)\right]^{2} \\
&+\sum_{L}\left[a_{0}\mathcal{F}_{+}^{\Sigma_{L}^{\prime\prime}}\left(q\right)+a_{1}\left(1+\delta^{\prime\prime}\left(\mathbf{q}^{2}\right)\right) \mathcal{F}_{-}^{\Sigma_{L}^{\prime\prime}}\left(q\right)\right]^{2},
\end{aligned}
\end{equation}
where the labels 0($+$) and 1($-$) distinguish isoscalar and isovector contributions, respectively, while the coupling constants $a_0$ and $a_1$ are form factors encoding information from the particle and hadronic sectors~\cite{PhysRevD.86.103511,PhysRevD.88.083516,*PhysRevD.89.029901,PhysRevD.102.074018}. 
The terms $\delta^{\prime}\left(\mathbf{q}^{2}\right)$ and $\delta^{\prime\prime}\left(\mathbf{q}^{2}\right)$ encode physics beyond leading spin-dependent coupling, combining effects of radius corrections, pseudoscalar form factors, and 2BCs:
\begin{equation}
\label{delta}
\begin{aligned}
\delta^{\prime}\left(\mathbf{q}^{2}\right) &=-\frac{\mathbf{q}^{2}\left\langle r_{A}^{2}\right\rangle}{6}+\delta a\left(\mathbf{q}^{2}\right) \\
\delta^{\prime \prime}\left(\mathbf{q}^{2}\right) &=-\frac{g_{\pi N N} F_{\pi}}{g_{A} m_{N}} \frac{\mathbf{q}^{2}}{\mathbf{q}^{2}+M_{\pi}^{2}}+\delta a\left(\mathbf{q}^{2}\right)+\delta a^{P}\left(\mathbf{q}^{2}\right),
\end{aligned}
\end{equation}
where  $\left\langle r_{A}^{2}\right\rangle$ is the axial radius, taken to be 0.46(16) fm$^2$ from a global analysis of muon capture and neutrino scattering~\cite{Hill_2018}. 
For the axial-vector coupling constant $g_A$, pion-decay constant $F_{\pi}$, $\pi N$ coupling constant $g_{\pi N N}$, we take values consistent with chiral nuclear forces. 

The 2BC contributions $\delta a\left(\mathbf{q}^{2}\right)$ and $\delta a^{P}\left(\mathbf{q}^{2}\right)$ are approximated via  normal-ordering with respect to a Fermi gas reference state with density $\rho$~\cite{PhysRevD.102.074018}, similar to $\beta$ and $\beta \beta$ decay studies \cite{PhysRevLett.107.062501,Gysbers2019}. 
To assess the accuracy of this approximation, we note that the $\delta a$ term that remains at $q=0$~MeV in Eq.(\ref{delta}) is the same as for Gamow-Teller transitions, where full 2BCs have been calculated. 
Since the full 2BC result lies within the range obtained by normal ordering with $\rho=0.08\ldots 0.12$ fm$^{-3}$~\cite{Gysbers2019}, we at least expect this to be reliable at low-momentum transfer, and similar investigations are underway for heavier nuclei.
As in Refs.~\cite{PhysRevD.88.083516,*PhysRevD.89.029901,PhysRevD.102.074018}, we take $\rho=0.08 \ldots 0.12$~fm$^{-3}$ and include all pion-exchange, pion-pole, and contact terms derived in Ref.~\cite{PhysRevD.102.074018}. 
For consistency we take the values for coupling constants and LECs (e.g., $c_1$, $c_3$, $c_4$ and $c_D$) that appear in chiral currents to be the same as in the particular chiral nucleon-nucleon (NN) interactions used in this work \cite{PhysRevC.83.031301,PhysRevC.93.011302,Gysbers2019,Hofe15RS,PhysRevC.102.054301,Siem16pn}. 
One should use 2BCs consistent with the chiral order of nuclear forces, but when there is a mismatch (e.g., the N$^4$LO+3N$_{\rm lnl}$ interaction), higher-order effects can be estimated by shifting LECs~\cite{Hofe15RS,HOFERICHTER20161}. We find such effects to be 10\% for structure factors.
Since we also consider a $\rm{\Delta}$-full interaction, we consistently include the 2BCs due to the excitation of a nucleon into a $\rm{\Delta}$ via pion exchange~\cite{PhysRevC.102.025501}.
Finally, we use a value of $c_6=5.01(1.06)$ GeV$^{-1}$ adopted from Table V in Ref.~\cite{PhysRevD.102.074018}, including uncertainty from chiral loops correction. 
See Supplemental Material for further details.

The structure factor $S_{A}(\mathbf{q})$ can be decomposed into isoscalar/isovector components $S_{00}$, $S_{01}$ and $S_{11}$ through
\begin{equation}
S_{A}(\mathbf{q})=a_{0}^{2} S_{00}(\mathbf{q})+a_{0} a_{1} S_{01}(\mathbf{q})+a_{1}^{2} S_{11}(\mathbf{q}).
\end{equation}
However, it is common in the literature to use the ``proton" ($S_{\rm p}$) and ``neutron" ($S_{\rm n}$) factors, defined by the couplings $a_0=a_1=1$ and $a_0=-a_1=1$, respectively. 
Here we focus on $S_{\rm p}$ and $S_{\rm n}$, but the $S_{00}$, $S_{01}$ and $S_{11}$ can be obtained from our public Jupyter notebook \cite{baishan_ipynb}.

To investigate uncertainties due to nuclear Hamiltonians, we start from three established chiral interactions and corresponding currents.
Specifically, we use 1.8/2.0(EM) \cite{PhysRevC.83.031301,Simo17SatFinNuc}, which reproduces ground-state energies to the heavy-mass region (while underpredicting charge radii)~\cite{Morr18Tin,Holt21Drip}; the  N$^4$LO+3N$_{\rm lnl}$ interaction of Refs.~\cite{Leis18Ti,Soma20LNL}; and the newly developed $\rm{\Delta NNLO_{GO}}$(394)~\cite{PhysRevC.102.054301} which includes explicit $\rm{\Delta}$-isobar degrees of freedom and optimizes NN and 3N forces simultaneously at the N$^2$LO level. 

For the nuclear calculation, we begin in a harmonic oscillator (HO) basis with frequency $\hbar \omega$ then transform the Hamiltonian and nuclear response operators to the Hartree-Fock (HF) or NAT basis constructed from the perturbatively improved one-body density matrix  \cite{PhysRevC.99.034321,PhysRevC.102.051303,PhysRevC.103.014321}. We take $e=2n+l \leq e_{\rm max}^{\rm HF/NAT}$,  with an additional truncation $e_1+e_2+e_3 \leq E_{\rm 3max}$ imposed for 3N  matrix elements. 
Finally, we use the Magnus formulation of the VS-IMSRG \cite{PhysRevLett.106.222502,PhysRevLett.113.142501,HERGERT2016165,PhysRevLett.118.032502,PhysRevC.102.034320,PhysRevC.92.034331} in a smaller subspace $e_{\rm max} \leq e_{\rm max}^{\rm HF/NAT}$ to decouple a valence-space Hamiltonian, using ensemble normal ordering to approximately capture 3N forces between valence nucleons~\cite{PhysRevLett.118.032502}. 
The operators for the nuclear response are consistently transformed to produce effective valence-space operators, truncated at the two-body level, the IMSRG(2) approximation. 
We take the same valence spaces used in LSSM calculations: $sd$ shell for $^{19}$F, $^{23}$Na, $^{27}$Al and $^{29}$Si; \{$p_{3/2}$, 0$f_{5/2}$, 1$p_{1/2}$, 0$g_{9/2}$\} orbits above a $^{56}$Ni core for $^{73}$Ge, and \{0$g_{7/2}$, 1$d_{5/2}$, 1$d_{3/2}$, 2$s_{1/2}$, 0$h_{11/2}$\} orbitals above a $^{100}$Sn core for $^{127}$I and $^{129,131}$Xe and perform exact diagonalizations with the KSHELL code~\cite{SHIMIZU2019372}. 

\begin{figure}[t]
\setlength{\abovecaptionskip}{0pt}
\setlength{\belowcaptionskip}{0pt}
\includegraphics[scale=0.54]{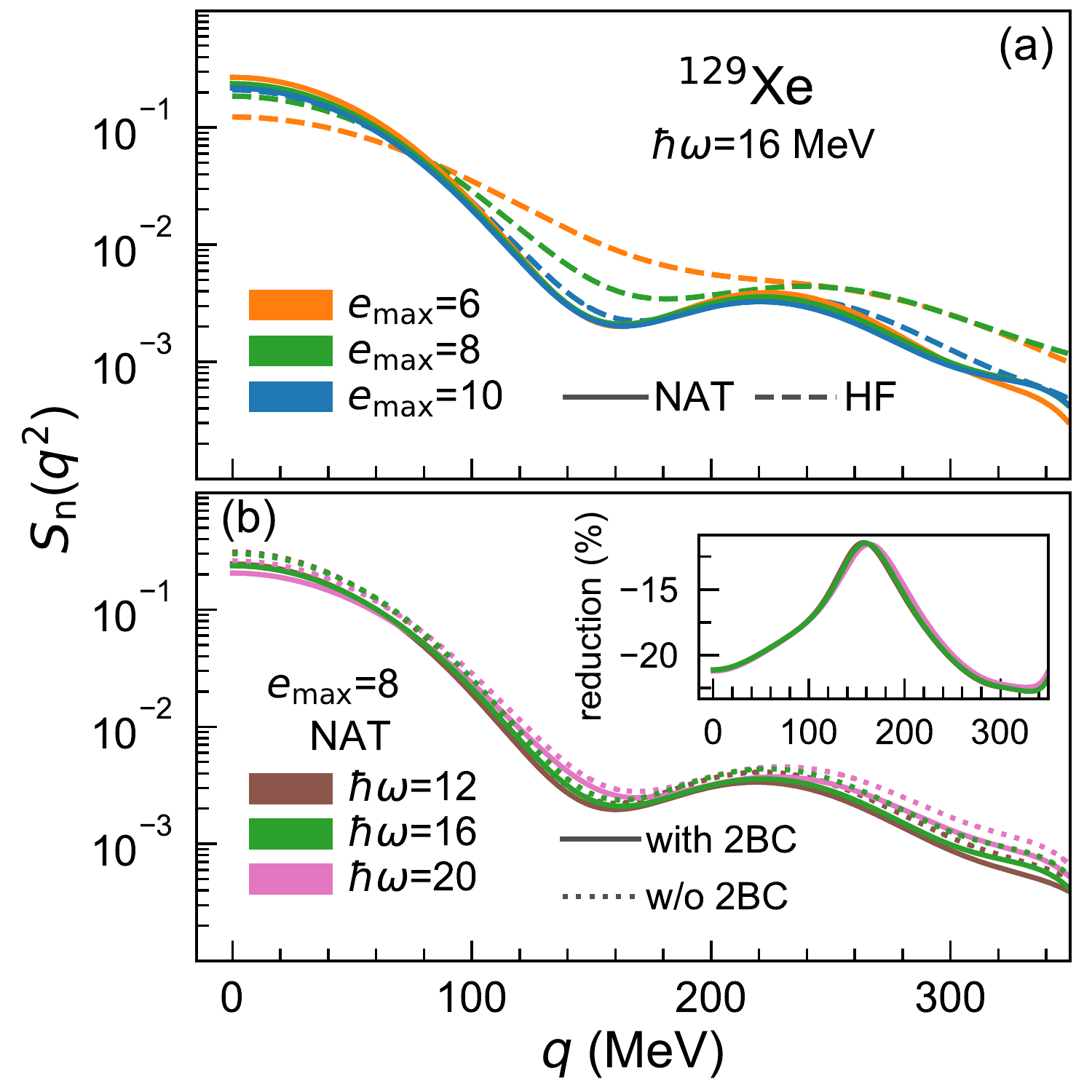}
\caption{\label{Xe129_convergence_Sn} 
(a) Convergence of the $^{129}$Xe structure factors $S_{\rm n}$ in the HF and NAT bases with the 1.8/2.0(EM) interaction.
(b) $\hbar \omega$ dependence of the results, where solid and dotted lines give calculations with and without 2BCs, respectively. The inset in panel (b) shows reduction rate of $S_{\rm n}$ caused by 2BCs.}
\end{figure}

Ab initio calculations of structure factors require considerable computational resources. 
In the present study, each nuclear response is calculated at 15 momentum transfer points, each requiring its own IMSRG transformation.
Furthermore $\mathcal{F}_{\pm}^{\Sigma_{L}^{\prime}}\left(q\right)$ and $\mathcal{F}_{\pm}^{\Sigma_{L}^{\prime\prime}}\left(q\right)$ are sums over the possible multipoles $L$ (e.g., $L$=1, 3 and 5 in $^{127}$I case) and the cost of the IMSRG transformation grows with $L$. 
Therefore convergence at relatively low $e_{\rm max}$ is necessary for such computations to be tractable in heavy nuclei.

In Fig.~\ref{Xe129_convergence_Sn}(a), we illustrate convergence of $S_{\rm n}$ within the HF and NAT bases for $^{129}$Xe, where $S_{\rm p}$ and all other studied nuclei follow a similar pattern.
For HF, we take the conventional $e_{\rm max}^{\rm HF}=e_{\rm max}$ truncation, while we use $e_{\rm max}^{\rm NAT}$=14 to construct the NAT basis, then perform the VS-IMSRG calculation in the smaller $e_{\rm max}$ $(< e_{\rm max}^{\rm NAT})$ subspace noted in the figure. 
While we find a rather slow convergence within the HF basis, in the NAT basis, there is almost no change from \mbox{$e_{\rm max}=8-10$}, demonstrating convergence is obtained by \mbox{$e_{\rm max}=8$}. 
In Fig.~\ref{Xe129_convergence_Sn}(b), we further confirm convergence at \mbox{$e_{\rm max}=8$} through variation of the basis parameter $\hbar \omega = 12-20$~MeV, with results nearly independent of this choice. 
We also note that the smallest $\hbar\omega$ gives the most rapid convergence, consistent with other studies of electroweak observables in heavy nuclei~\cite{baishan}, suggesting a choice of $\hbar\omega=12$~MeV is preferable in the $^{132}$Sn region. 
Finally in panel (b), we also illustrate the effects of including 2BCs, that result in a $\sim 20\%$ suppression (q $\lesssim$100~MeV region) of the structure factor similar to the previous LSSM works~\cite{PhysRevD.88.083516,*PhysRevD.89.029901}.

Based on convergence trends, we take the following prescription for our final structure factor calculations. 
For nuclei in the $sd$ shell and $^{73}$Ge, we use the HF basis with $e_{\rm max}^{\rm HF}=e_{\rm max}=10$, while for the heavy nuclei $^{129}$I and $^{129,131}$Xe, we use the NAT basis with $e_{\rm max}^{\rm NAT}=14$ and $e_{\rm max}=8$. 
In all cases we use $\hbar\omega=16~$MeV and $E_{\rm 3max}=22$, which, as illustrated in the Supplemental Material, is sufficient to obtain converged structure factors. 
While a large $E_{\rm 3max}$ is necessary to obtain converged ground-state energies of heavy nuclei~\cite{takayuki}, we note structure factors are largely converged by $E_{\rm 3max}=18$.

\begin{figure}[t]
\setlength{\abovecaptionskip}{0pt}
\setlength{\belowcaptionskip}{0pt}
\includegraphics[scale=0.42]{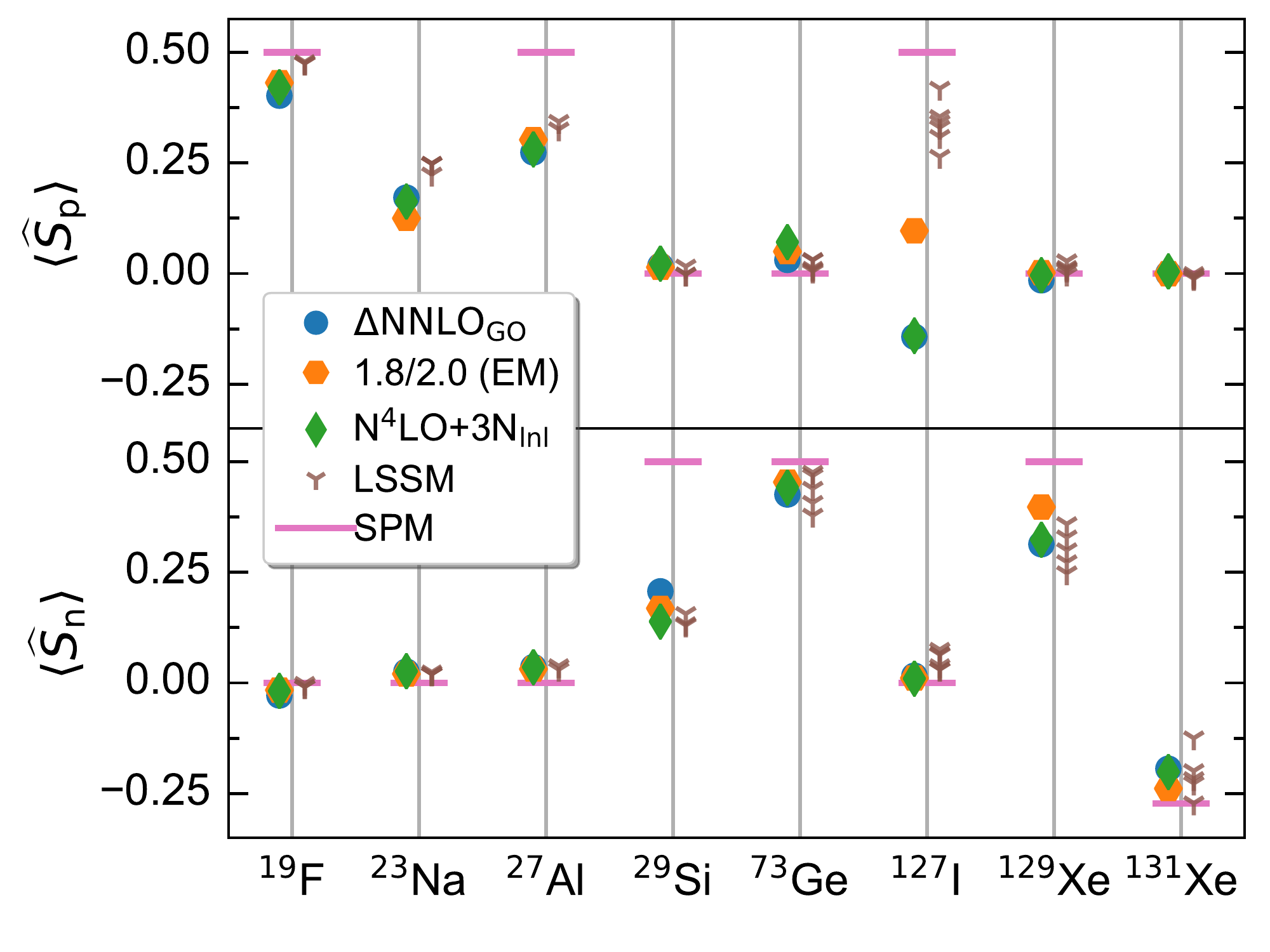}
\caption{\label{Sigma} Ab initio results of spin expectation for proton $\widehat{S}_{\rm p}$ and neutron $\widehat{S}_{\rm n}$, compared to LSSM calculations~\cite{PhysRevD.88.083516,*PhysRevD.89.029901} and single-particle model (SPM) estimates of Ref.~\cite{doi:10.1142/S0218301392000023}.}
\end{figure}

At $q=0$~MeV, the structure factor is given by~\cite{doi:10.1142/S0218301392000023}:
\begin{equation}
\begin{array}{l}
S_{A}(0)=\dfrac{(J+1)(2J+1)}{4 \pi J}\left| \left(a_{0}+a_{1}^{\prime}\right)\left\langle\widehat{S}_{\rm p}\right\rangle+\left(a_{0}-a_{1}^{\prime}\right)\left\langle\widehat{S}_{\rm n}\right\rangle \right|^{2}
\end{array}
\end{equation}
where $\widehat{S}_{\rm p}$ and $\widehat{S}_{\rm n}$ denote the total spin for protons and neutrons, respectively, and $a_{1}^{\prime}=a_1(1+ \delta a(0))$ includes the effects from chiral 2BCs, namely $\delta a(0)$.
In Fig.~\ref{Sigma} we show the calculated spin expectation values compared with the naive single-particle model (SPM) estimate of Ref.~\cite{doi:10.1142/S0218301392000023} and LSSM results with different phenomenological Hamiltonians~\cite{PhysRevD.88.083516,*PhysRevD.89.029901}. 
Generally the VS-IMSRG gives similar results independent of starting Hamiltonian and consistent with the LSSM, though the uncertainties in heavy nuclei, as estimated by the spread in values, are somewhat smaller than the those from the LSSM. 
As expected for odd-mass nuclei, there is a clear hierarchy of the spin expectation values with either $|\langle \widehat{S}_{\rm p} \rangle |\gg|\langle \widehat{S}_{\rm n} \rangle |$ or $|\langle \widehat{S}_{\rm n} \rangle |\gg|\langle \widehat{S}_{\rm p} \rangle |$, depending on the unpaired nucleon.
We note that the ab initio result for $^{127}$I gives the biggest difference compared with SPM and LSSM. 
This is because the wave functions obtained with the $\Delta$NNLO$_{\rm GO}$ and N$^{4}$LO+3N$_{\rm lnl}$ interactions are dominated by a $(0g_{7/2})^3$ proton configuration,
while the 1.8/2.0(EM) wave function has a larger contribution from a $\{(0g_{7/2})^2(1d_{5/2})^1\}$ configuration, leading to the opposite spin.

\begin{figure*}
\setlength{\abovecaptionskip}{0pt}
\setlength{\belowcaptionskip}{0pt}
\includegraphics[scale=0.70]{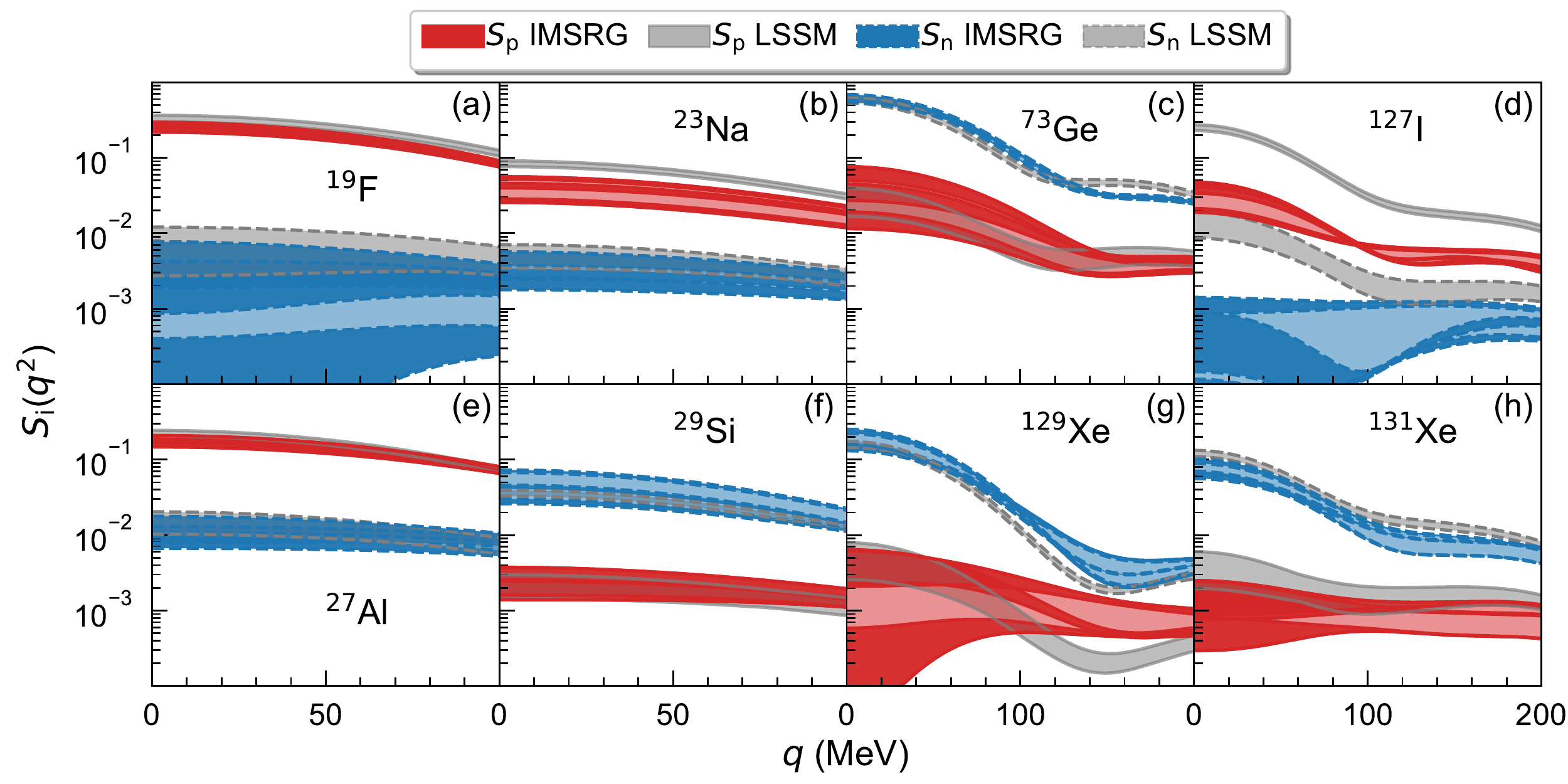}
\caption{\label{all-shell} Ab initio structure factors $S_{\rm p}$ and $S_{\rm n}$ for $^{19}$F, $^{23}$Na, $^{27}$Al, $^{29}$Si, $^{73}$Ge, $^{127}$I, and $^{129,131}$Xe as a function of momentum transfer $q$. 
The grey bands indicate LSSM calculations, where results in $^{129,131}$Xe are obtained from Ref~\cite{PhysRevD.102.074018}. For other nuclei, by taking fit coefficients of nuclear response functions in Ref.~\cite{PhysRevD.102.074018}, we give LSSM results including 2BCs with all pion-exchange, pion-pole and contact terms. The VS-IMSRG bands indicate the spread in results from different interactions (lighter bands) and uncertainties in 2BCs (darker bands), while LSSM bands are from 2BC uncertainties only. We also give the structure factors in polar coordinates  in the Supplemental Material to facilitate experimental comparisons.}
\end{figure*}

To generate our final results, we note that in the case of the HO basis, the matrix elements of $\mathcal{F}_{\pm}^{\Sigma_{L}^{\prime}}\left(q\right)$ and $\mathcal{F}_{\pm}^{\Sigma_{L}^{\prime\prime}}\left(q\right)$ can be evaluated analytically, yielding forms $\sim e^{-\frac{u}{2}}p(u)$ where $p(u)$ is a polynomial in the dimensionless variable $u=q^2 b^2/2$ and $b=\sqrt{\hbar/(m \omega)}$. 
Although ab initio many-body wavefunctions are composed of a mixture of HO configurations, the final $\mathcal{F}_{\pm}^{\Sigma_{L}^{\prime}}\left(q\right)$ and $\mathcal{F}_{\pm}^{\Sigma_{L}^{\prime\prime}}\left(q\right)$ still follow the $e^{-\frac{u}{2}}p(u)$ forms. 
By fitting the results at the 15 calculated $q$ points, we obtain smooth curves for structure factors as a function of $q$. 
Our Jupyter notebook gives the fit coefficients, so structure factors can then be evaluated at any $q$ point.

Figure~\ref{all-shell} shows the final structure factors for all spin-dependent detector nuclei, compared to those of phenomenological LSSM calculations \cite{PhysRevD.102.074018}.
The overall VS-IMSRG error band takes into account the uncertainty due to the input nuclear interactions as well as the parameter $\rho=0.08 \ldots 0.12$ fm$^{-3}$ in the treatment of 2BCs, while the LSSM uncertainty bands originate from the LECs $c_3$, $c_4$ and the density $\rho$ range. 
We note the uncertainty from higher-order ($>Q^3$) one-body and two-body currents is not included in Fig.~\ref{all-shell}.
We first see that the structure factors $S_{\rm p}$ and $S_{\rm n}$ have a clear hierarchy for all calculated nuclei, which is consistent with the spin expectation values. 
The dominant structure factors are in overall agreement with the LSSM predictions, while for the non-dominant factors, ab initio calculations generally have a larger uncertainty at q $\lesssim$100~MeV, primarily resulting from the 2BCs from $\Delta$NNLO$_{\rm GO}$.
As shown in Figs.~\ref{rate_dominant} and \ref{rate_non-dominant} in the Supplemental Material, the largest differences with LSSM arise in $^{23}$Na and particularly $^{127}$I, where the minimum value of the dominant factor band is reduced by 66\% and 92\%, respectively ($^{29}$Si gives a 120\% enhancement, but the LSSM $S_{\rm n}$ in $^{29}$Si is smaller than $S_{\rm p}$ in $^{23}$Na or $^{127}$I).

Due to the corrections shown in Eq.~\eqref{delta}, the dominance of either structure factor $S_{\rm p}(\mathbf{q}^{2})$ or $S_{\rm n}(\mathbf{q}^{2})$ becomes somewhat less clear.
The earlier LSSM calculations demonstrated that the $\delta a(\mathbf{q}^{2})$ and $\delta a^{P}(\mathbf{q}^{2})$ give the dominant corrections of structure factors in the small $q$ region $\lesssim 100$ MeV~\cite{PhysRevD.88.083516,*PhysRevD.89.029901}.
Similar to previous works, our calculations show that 2BCs suppress the dominant structure factor by 10\%$-$35\% at low $q$.
Also, enhancement of the non-dominant structure factor is observed for the 1.8/2.0(EM) and N$^4$LO+3N$_{\rm lnl}$ interactions.
Although 2BCs from $\rm{\Delta NNLO_{GO}}$ tend to increase the non-dominant factor significantly, we find a reduction in $^{19}$F and $^{127}$I.
This is a primary source of the large uncertainty bands shown in Fig.~\ref{all-shell}.
Since the relevant LECs employed in $\Delta$-less and $\Delta$-full EFTs are quite different (see Supplemental Material), this is somewhat expected.  
We note that the $\Delta$ excitation effects are compensated by the relatively large LECs in the $\Delta$-less chiral EFT, and the final corrections $\delta a(\mathbf{q}^{2})$ and $\delta a^{P}(\mathbf{q}^{2})$ are comparable in both EFTs, except for the $^{19}$F and $^{127}$I cases.
Moreover, we find that $\delta a^{P}(\mathbf{q}^{2})$ is overall suppressed by the contact-$c_{D}$ and pion-pole terms, leading to an always negative longitudinal correction $\delta^{\prime \prime}\left(\mathbf{q}^{2}\right)$ in Eq.~\eqref{delta}. 

Finally, while our focus has been on structure factors, in the Supplemental Material we show ground-state energies and spectra for the nuclei studied. 
We note the at times poor agreement with experiment, particularly for heavy nuclei (where deformation is likely not adequately captured), but it is not clear to what extent this appreciably impacts DM scattering, since, e.g., spin expectation values are nevertheless very similar to those of the LSSM, which well reproduces all spectra. 
To this end we have performed a preliminary systematic exploration using a number of initial NN+3N forces spanning a vast space of LECs~\cite{Hu21208Pb} and find little to no correlation between spectroscopic quality and spin expectation values.

In summary we have used the VS-IMSRG to compute ab initio spin-dependent structure factors for all nuclei currently used in direct detection searches. Using three different NN+3N interactions from chiral EFT with consistent 2BCs, we obtain convergence even in heavy nuclei using the NAT basis with $e_{\rm max}=8$ and $E_{\rm 3max}=22$. 
Overall results are consistent with LSSM calculations, but notable discrepancies at low momentum transfer can be seen in several cases, highlighting the need for further improvements.
While the work presented here is a promising first step, additional sources of uncertainty remain to be explored, such as the IMSRG(3) truncation~\cite{Heinz21IMSRG3} and direct inclusion of 2BCs without the normal-ordering approximation. 
Furthermore, non-relativistic EFT~\cite{Fitzpatrick_2013,PhysRevD.95.103011} is a complementary approach to tackle the physics of WIMPs-nucleon interactions, and has already been implemented within the VS-IMSRG to provide future comparisons with the chiral EFT results. 
Finally, work is currently in progress to extend these calculations to nuclear structure factors or nuclear responses for spin-independent WIMP-nucleus scattering, inelastic WIMP scattering and coherent elastic neutrino-nucleus scattering.
All results presented here are publicly available in a Jupyter notebook (see Supplemental Material).  

\begin{acknowledgments}
We thank M. Hoferichter, J. Men\'endez, A. Robinson, and A. Schwenk for insightful discussions, benchmarking, and comments on the manuscript.
TRIUMF receives funding via a contribution through the National Research Council of Canada.
This work was further supported by NSERC under grants SAPIN-2018-00027 and RGPAS-2018-522453, the Arthur B. McDonald Canadian Astroparticle Physics Research Institute, the Canadian Institute for Nuclear Physics, the MITACS Globalink Research Internship program, and the US Department of Energy (DOE) under contract DE-FG02-97ER41014.
Computations were performed with an allocation of computing resources on Cedar at WestGrid and Compute Canada, and on the Oak Cluster at TRIUMF managed by the University of British Columbia department of Advanced Research Computing (ARC).
\end{acknowledgments}

\bibliography{references}

\section{\label{Supplementary}Supplemental Material}

\begin{figure*}
\setlength{\abovecaptionskip}{0pt}
\setlength{\belowcaptionskip}{0pt}
\includegraphics[scale=0.70]{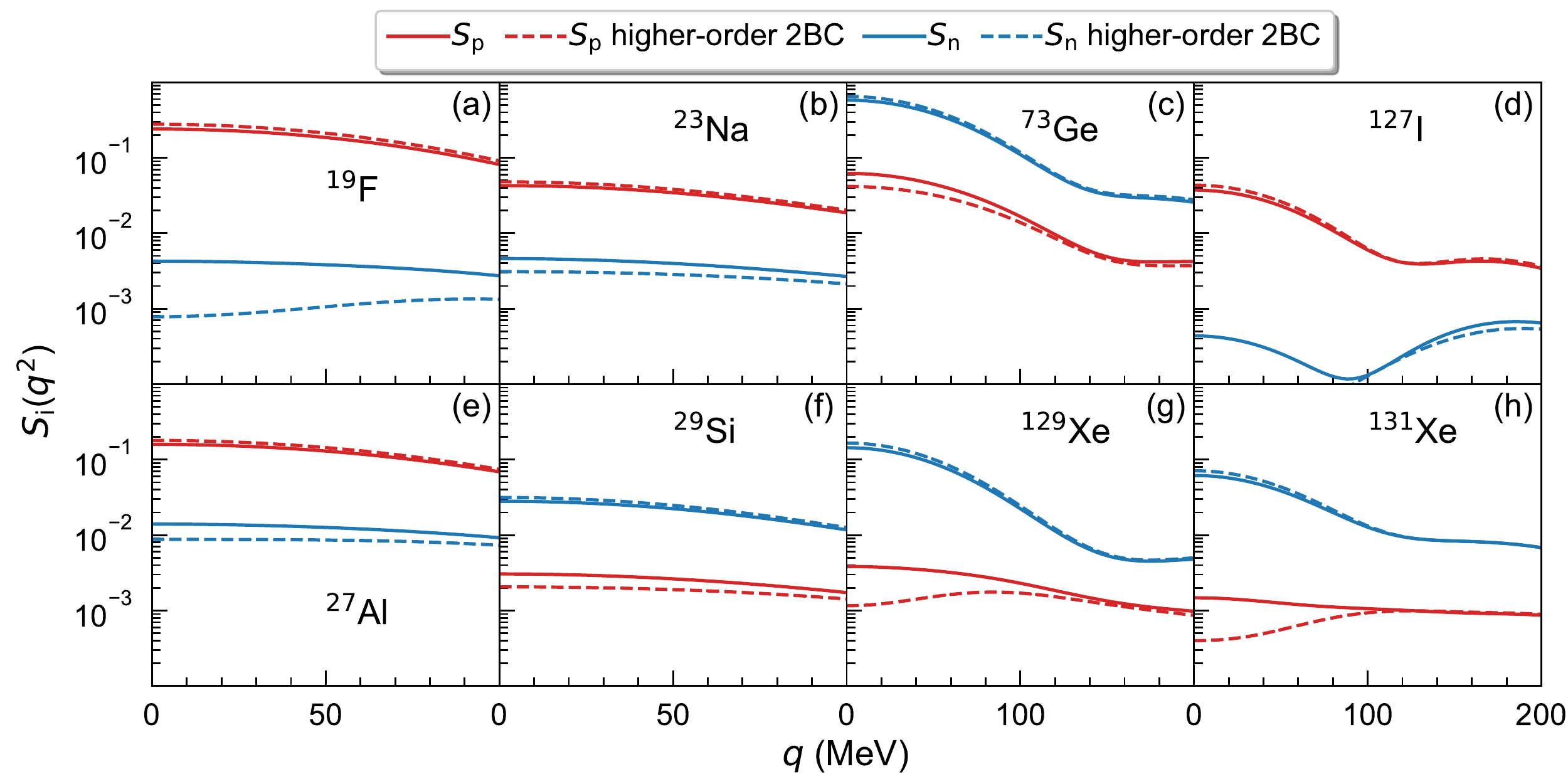}
\caption{\label{higher_order} Ab initio structure factors $S_{\rm p}$ and $S_{\rm n}$ calculated by tree-level two-body current (2BC), compared with results of adding partial corrections from higher-order 2BC. The higher-order effects are captured in the LECs, $c_1$, $c_3$ and $c_4$, see table~\ref{LECs} and text for the details about the LECs. The N$^4$LO+3N$_{\rm lnl}$ interaction and NAT basis with $e^{\rm NAT}_{\rm max}=14$ and $e_{\rm max}=8$ are used. We take the parameter $\rho$=0.10 fm$^{-3}$ in the treatment of 2BCs.}
\end{figure*}

Table~\ref{LECs} shows the LECs employed in the $\Delta$-less and $\Delta$-full chiral Hamiltonians used in this work.
For $\Delta$NNLO$_{\rm GO}$(394) a 2BC term for the excitation of a nucleon to a $\Delta$ by pion exchange arises at order $Q^2$ \cite{PhysRevC.98.044003,PhysRevC.102.025501}. 
The topology of this diagram is identical to the structure of the $\Delta$-less $\pi$-exchange 2BC term when using the resonance-saturation values for the $\pi$N LECs, $c_{3}^{\Delta}=-2 c_{4}^{\Delta}=\frac{4 h_{A}^{2}}{9 \delta}=-2.972246$ GeV$^{-1}$ \cite{PhysRevC.97.024332,PhysRevC.98.044003,PhysRevC.102.025501}. 
Therefore, we include the 2BCs due to $\Delta$ by shifting the LECs in row `$\Delta$NNLO$_{\rm GO}$' to row `$\Delta$NNLO$_{\rm GO}$*' of Table~\ref{LECs}.
For N$^4$LO+3N$_{\rm lnl}$, the NN sector is constructed to N$^4$LO, while the 3N interaction is at N$^2$LO. 
Since the two-pion exchange 3N force has the same structure at N$^2$LO, N$^3$LO and N$^4$LO, one can include N$^3$LO and N$^4$LO components of 3N forces in N$^2$LO by shifting the parameters of the two-pion-exchange 3N forces ($c_1$, $c_3$ and $c_4$) \cite{PhysRevC.77.064004,PhysRevC.96.024004,PhysRevC.85.054006} with respect to their values in the NN potential~\cite{Hofe15RS,HOFERICHTER20161}. 
Since the axial-vector 2BC is matched to the 3N force \cite{KREBS2017317}, we estimate higher-order 2BCs by shifting the effective LECs, $c_1$, $c_3$ and $c_4$, from row `N$^4$LO+3N$_{\rm lnl}$' to row `N$^4$LO+3N$_{\rm lnl}$*' of Table~\ref{LECs}.
Figure~\ref{higher_order} shows the structure factor calculated using LECs of `N$^4$LO+3N$_{\rm lnl}$' and `N$^4$LO+3N$_{\rm lnl}$*'; we find differences at the level of 10\%.

In Table~\ref{Sigma_value} we list the spin expectation values of proton $\widehat{S}_{p}$ and neutron $\widehat{S}_{n}$; large-scale shell model (LSSM) results from Ref. \cite{PhysRevD.88.083516,*PhysRevD.89.029901} are also shown for comparison. 

\begin{figure*}
\setlength{\abovecaptionskip}{0pt}
\setlength{\belowcaptionskip}{0pt}
\includegraphics[scale=0.70]{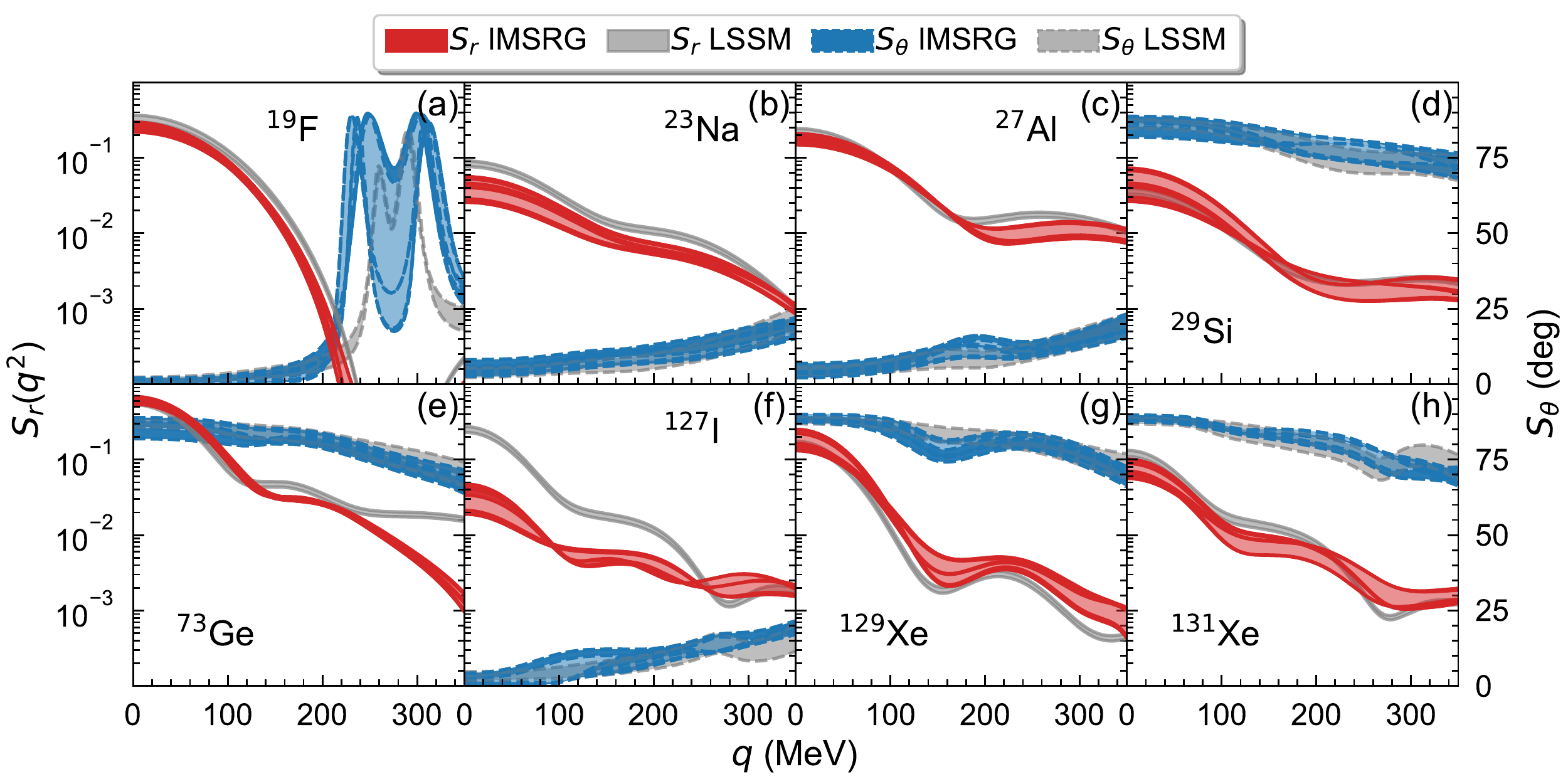}
\caption{\label{all-shell_polar} Ab initio structure factors $S_{\rm p}$ and $S_{\rm n}$ for $^{19}$F, $^{23}$Na, $^{27}$Al, $^{29}$Si, $^{73}$Ge, $^{127}$I, and $^{129,131}$Xe as a function of momentum transfer $q$. 
The grey bands indicate LSSM calculations, where results in $^{129,131}$Xe are obtained from Ref~\cite{PhysRevD.102.074018}. 
For other nuclei, by taking fit coefficients of nuclear response functions in Ref.~\cite{PhysRevD.102.074018}, we give LSSM results including 2BCs with all pion-exchange, pion-pole and contact terms. 
The VS-IMSRG bands indicate the spread in results from different interactions (lighter bands) and uncertainties in 2BCs (darker bands), while LSSM bands are from 2BC uncertainties only.  
Note that $S_r$ displayed on the logarithm scale, and $S_{\theta}$ is shown with the standard scale.}
\end{figure*}

Figure~\ref{all-shell_polar} illustrates structure factors $S_{\rm p}$ and $S_{\rm n}$ in polar coordinates by defining $
S = S_{\rm p} + i S_{\rm n} = S_re^{iS_{\theta}}$,
where $S_r = |S|= \sqrt{S^2_{\rm p} + S^2_{\rm n}}$ is the modulus of complex $S$, and $S_{\theta} = {\rm Arg}(S)$ is the argument. 
The LSSM $S_{\rm p}$ and $S_{\rm n}$ values are obtained from Ref~\cite{PhysRevD.102.074018}. 

\begin{table*}
\centering
\caption{\label{LECs}
Values of the low-energy coupling (LEC) $c_1$, $c_3$, $c_4$ and $c_{D}$ for the chiral NN+3N forces used in this work. The N$^4$LO+3N$_{\rm lnl}$ gives the LEC used in the NN interaction, while N$^4$LO+3N$_{\rm lnl}$* shows 3N LEC. See the text for the details about $\rm{\Delta NNLO_{GO}}$(394) and $\rm{\Delta NNLO_{GO}}$(394)*.}
\begin{tabular*}{120mm}{c@{\extracolsep{\fill}}ccccc}
\hline
\hline
Interaction & $c_1$ & $c_3$ &  $c_4$  & $c_{D}$ \\
\hline
1.8/2.0(EM)~\cite{PhysRevC.83.031301,PhysRevC.93.011302}  & -0.81 & -3.20  & 5.40  & 1.264 \\
N$^4$LO+3N$_{\rm lnl}$~\cite{Gysbers2019,Hofe15RS} & -1.10 & -5.54  & 4.17  & -1.800 \\
N$^4$LO+3N$_{\rm lnl}$* & -0.73 & -3.38  & 1.69  & -1.800 \\
$\rm{\Delta NNLO_{GO}}$(394)~\cite{PhysRevC.102.054301,Siem16pn}  & -0.74 & -0.65  & 0.96  & 0.081 \\
$\rm{\Delta NNLO_{GO}}$(394)*  & -0.74 & -3.62  & 2.45  & 0.081 \\
\hline
\hline
\end{tabular*}
\end{table*}

\begin{table*}
\centering
\caption{\label{Sigma_value}
Ab initio spin expectation for proton $\widehat{S}_{p}$ and neutron $\widehat{S}_{n}$ of $^{19}$F, $^{23}$Na, $^{27}$Al, $^{29}$Si, $^{73}$Ge, $^{127}$I and $^{129,131}$Xe, compared with the phenomenological LSSM results \cite{PhysRevD.88.083516,*PhysRevD.89.029901}.}
\begin{tabular*}{160mm}{c@{\extracolsep{\fill}}cccccccc}
\hline
\hline
\multirow{2}{*}{Nuclei} & \multicolumn{2}{c}{$\rm{\Delta NNLO_{GO}}$(394)} & \multicolumn{2}{c}{1.8/2.0(EM)} & \multicolumn{2}{c}{N$^4$LO+3N$_{\rm lnl}$} & \multicolumn{2}{c}{LSSM \cite{PhysRevD.88.083516,*PhysRevD.89.029901}}\\
\cline{2-3}
\cline{4-5}
\cline{6-7}
\cline{8-9}
 & $\widehat{S}_{p}$ & $\widehat{S}_{n}$ &  $\widehat{S}_{p}$  & $\widehat{S}_{n}$ & $\widehat{S}_{p}$ & $\widehat{S}_{n}$ & $\widehat{S}_{p}$ & $\widehat{S}_{n}$ \\
\hline
  $^{19}$F  &    0.4018 &   -0.0286 &          0.4310 &   -0.0163 &          0.4205 &   -0.0184 & 0.478 & -0.002 \\
  $^{23}$Na &    0.1714 &    0.0254 &          0.1248 &    0.0203 &          0.1624 &    0.0263 & 0.224 & 0.024 \\
  $^{27}$Al &    0.2736 &    0.0357 &          0.3022 &    0.0312 &          0.2808 &    0.0357 & 0.326 & 0.038 \\
  $^{29}$Si &    0.0177 &    0.2068 &          0.0141 &    0.1684 &          0.0227 &    0.1386 & 0.016 & 0.156 \\
  $^{73}$Ge &    0.0307 &    0.4257 &          0.0496 &    0.4536 &          0.0710 &    0.4415 & 0.031 & 0.439 \\
 $^{127}$I  &   -0.1429 &    0.0162 &          0.0965 &    0.0108 &         -0.1406 &    0.0097 & 0.342 & 0.031 \\
 $^{129}$Xe &   -0.0154 &    0.3133 &          0.0018 &    0.3973 &         -0.0042 &    0.3237 & 0.010 & 0.329 \\
 $^{131}$Xe &   -0.0001 &   -0.1940 &         -0.0005 &   -0.2387 &          0.0044 &   -0.2013 & -0.009 & -0.272 \\
\hline
\hline
\end{tabular*}
\end{table*}

\begin{figure}
\setlength{\abovecaptionskip}{0pt}
\setlength{\belowcaptionskip}{0pt}
\includegraphics[scale=0.56]{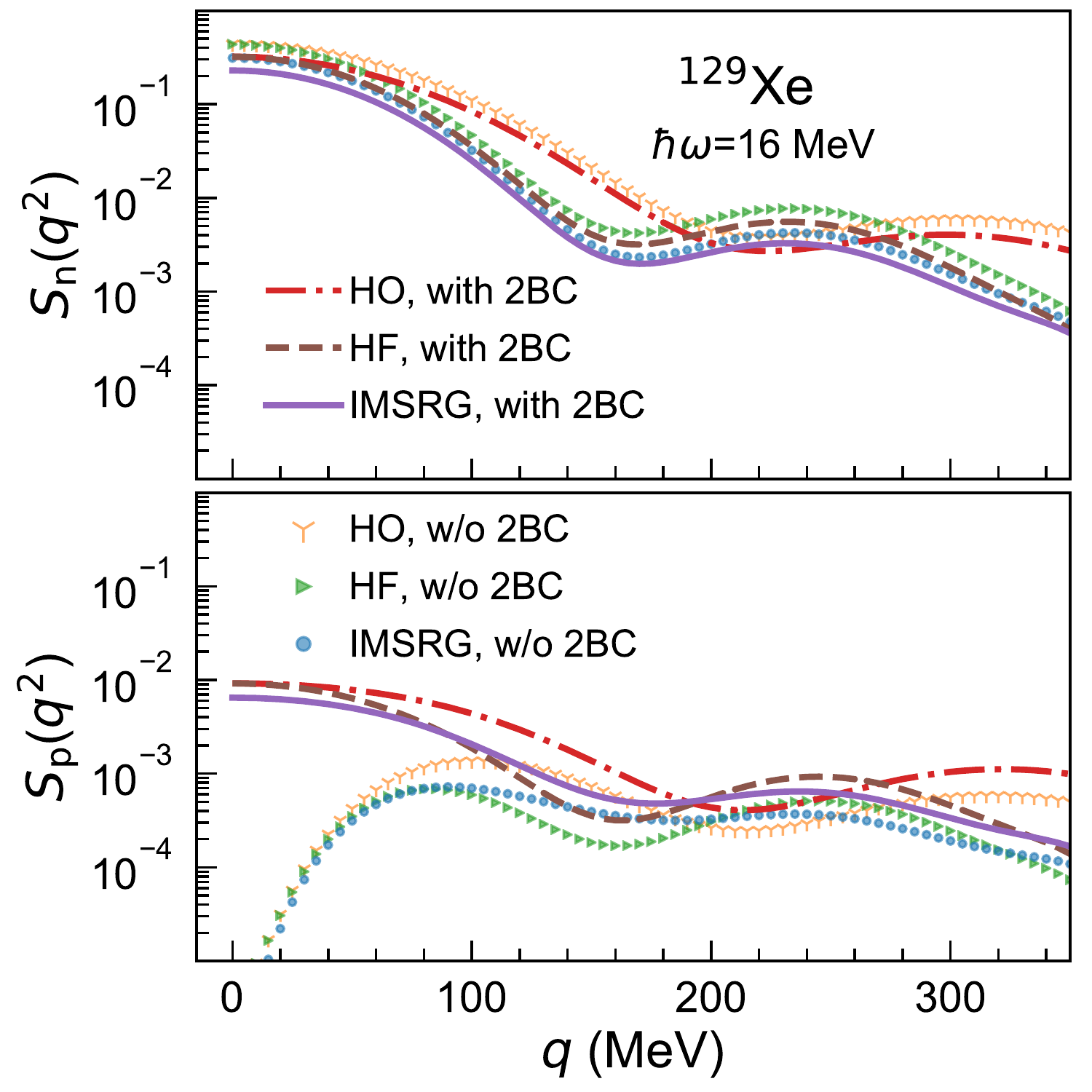}
\caption{\label{Xe129_IMSRG_HF} Structure factors $S_{\rm n}$ and $S_{\rm p}$ for $^{129}$Xe using the operator in the HO and HF bases with the IMSRG-evolved wavefunctions as well as the fully evolved IMSRG results (IMSRG), where the 1.8/2.0(EM) interaction and NAT basis with $e^{\rm NAT}_{\rm max}=14$ and $e_{\rm max}=8$ are used.}
\end{figure}

\begin{figure}
\setlength{\abovecaptionskip}{0pt}
\setlength{\belowcaptionskip}{0pt}
\includegraphics[scale=0.56]{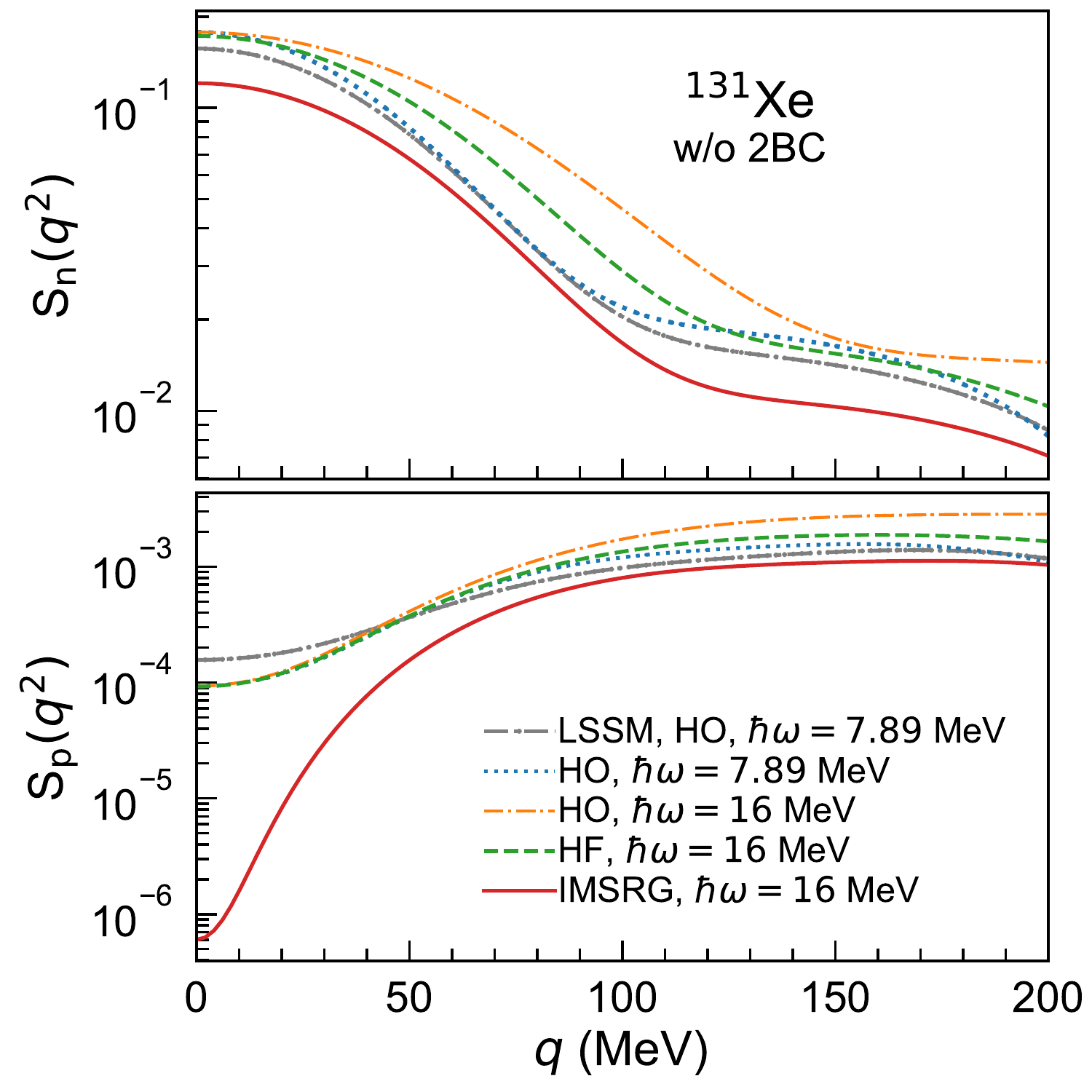}
\caption{\label{Xe131_IMSRG_HF} Structure factors $S_{\rm n}$ and $S_{\rm p}$ for $^{131}$Xe using the operator in the HO and HF bases with the IMSRG-evolved wavefunctions, as well as the fully evolved IMSRG results (IMSRG), using the 1.8/2.0(EM) interaction and NAT basis with $e^{\rm NAT}_{\rm max}=14$ and $e_{\rm max}=8$. We note LSSM calculations take $\hbar\omega=45A^{-1/3}-25A^{-2/3} \approx 7.89$ MeV. }
\end{figure}

\begin{figure}
\setlength{\abovecaptionskip}{0pt}
\setlength{\belowcaptionskip}{0pt}
\includegraphics[scale=0.56]{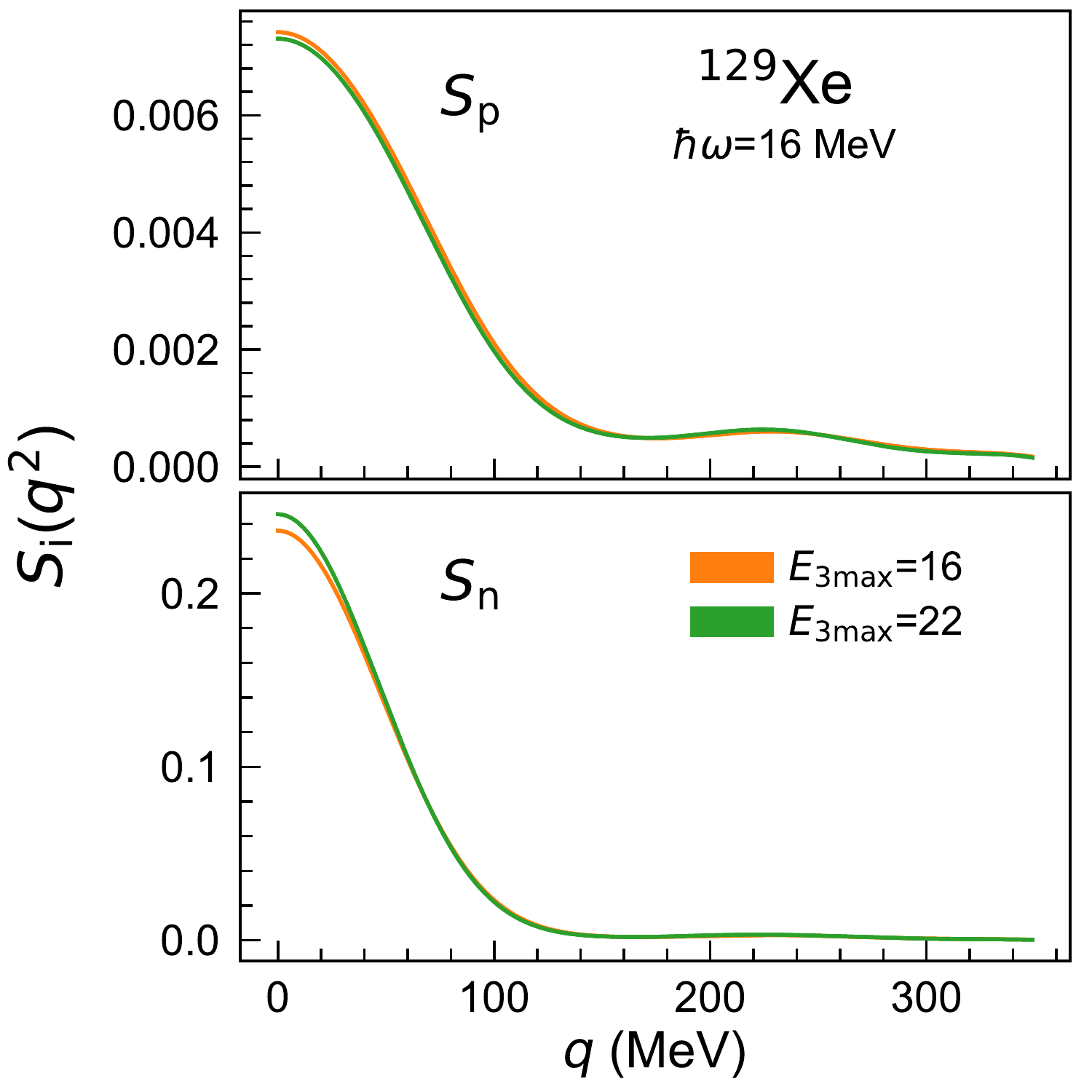}
\caption{\label{Xe129_convergence_E3max} Structure factors $S_{\rm n}$ and $S_{\rm p}$ $^{129}$Xe using different $E_{\rm 3max}$ truncations, where the 1.8/2.0(EM) interaction and NAT basis with $e^{\rm NAT}_{\rm max}=14$ and $e_{\rm max}=8$ are used.}
\end{figure}

In Figs.~\ref{Xe129_IMSRG_HF} and \ref{Xe131_IMSRG_HF} we show the $^{129,131}$Xe structure factor calculated by using the unevolved operator in both the harmonic oscillator (HO) and Hartree-Fock (HF) bases with (albeit inconsistent) VS-IMSRG wave functions, as well as the IMSRG-transformed operator consistent with the wave functions. 
The HF curves are closer to the final IMSRG results than HO calculations. Note that the LSSM approaches use unrenormalized bare operator within HO basis. 
$E_{\rm 3max}$ convergence of the $^{129}$Xe structure factors is given in Fig.~\ref{Xe129_convergence_E3max}.
$E_{\rm 3max}=16$ is commonly used in current nuclear structure calculations.  

\begin{figure}
\setlength{\abovecaptionskip}{0pt}
\setlength{\belowcaptionskip}{0pt}
\includegraphics[scale=0.4]{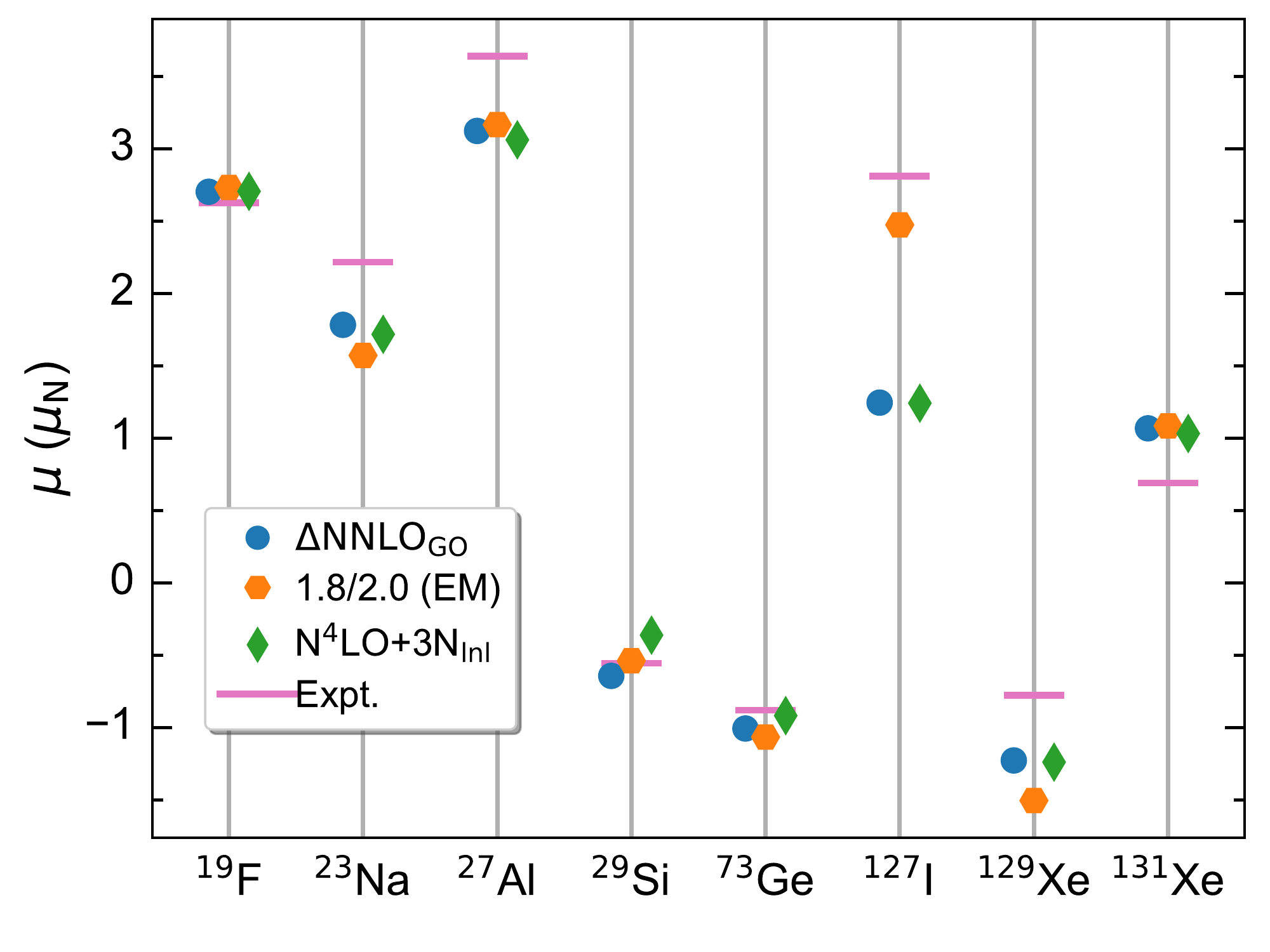}
\caption{\label{EM_moments} {Ab initio VS-IMSRG magnetic dipole moments, compared to experimental data \cite{STONE200575}, where effects of 2BCs are currently neglected.}}
\end{figure}

Figure~\ref{EM_moments} shows magnetic dipole moments $\mu$ of the target nuclei calculated from the VS-IMSRG using the chiral interactions discussed above; note 2BCs are not included.

\begin{figure*}
\setlength{\abovecaptionskip}{0pt}
\setlength{\belowcaptionskip}{0pt}
\includegraphics[scale=0.62]{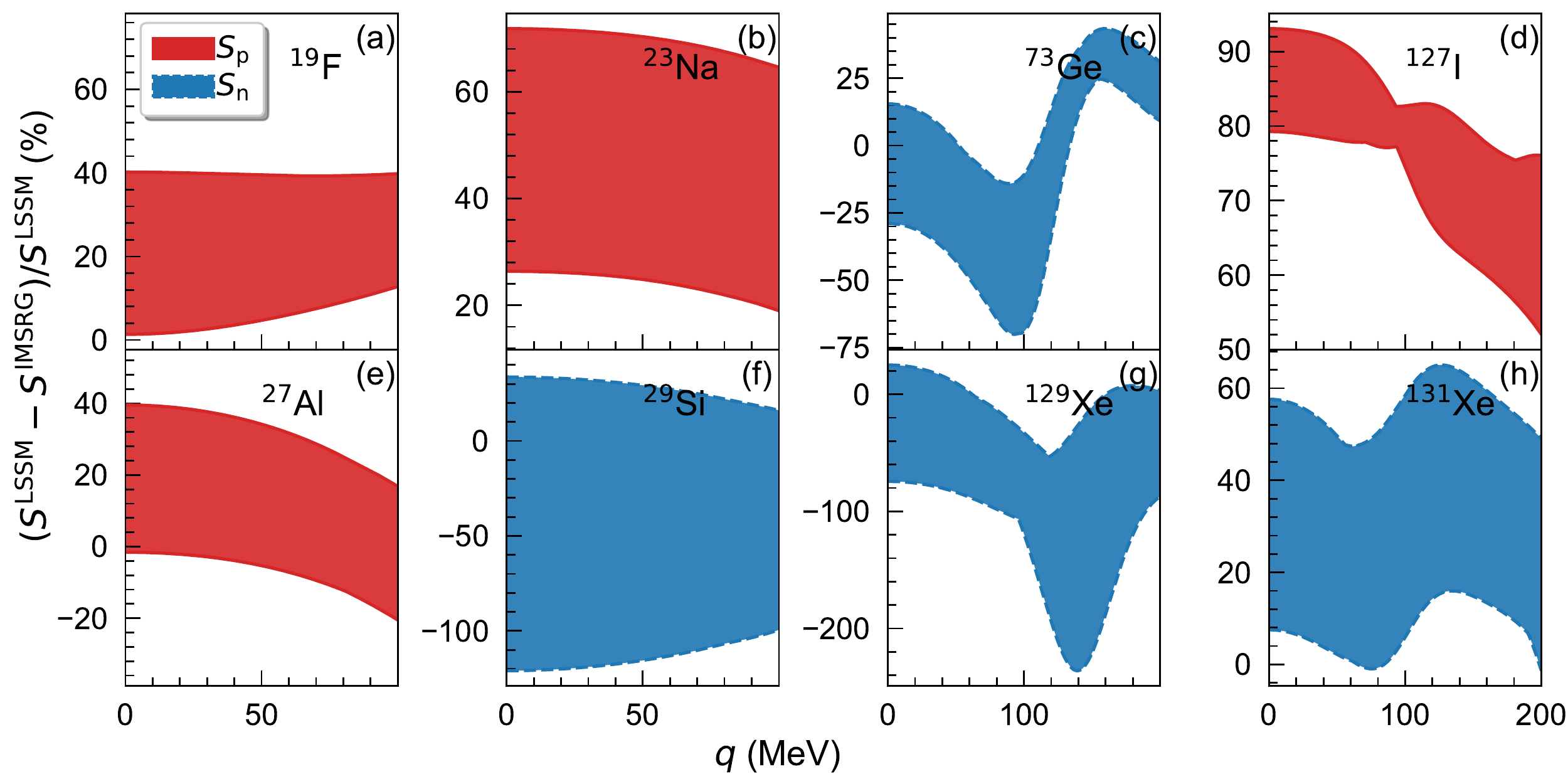}
\caption{\label{rate_dominant} 
Discrepancy of dominant structure factors between LSSM and VS-IMSRG. 
All calculations are same as in Fig.~\ref{all-shell}.}
\end{figure*}

\begin{figure*}
\setlength{\abovecaptionskip}{0pt}
\setlength{\belowcaptionskip}{0pt}
\includegraphics[scale=0.62]{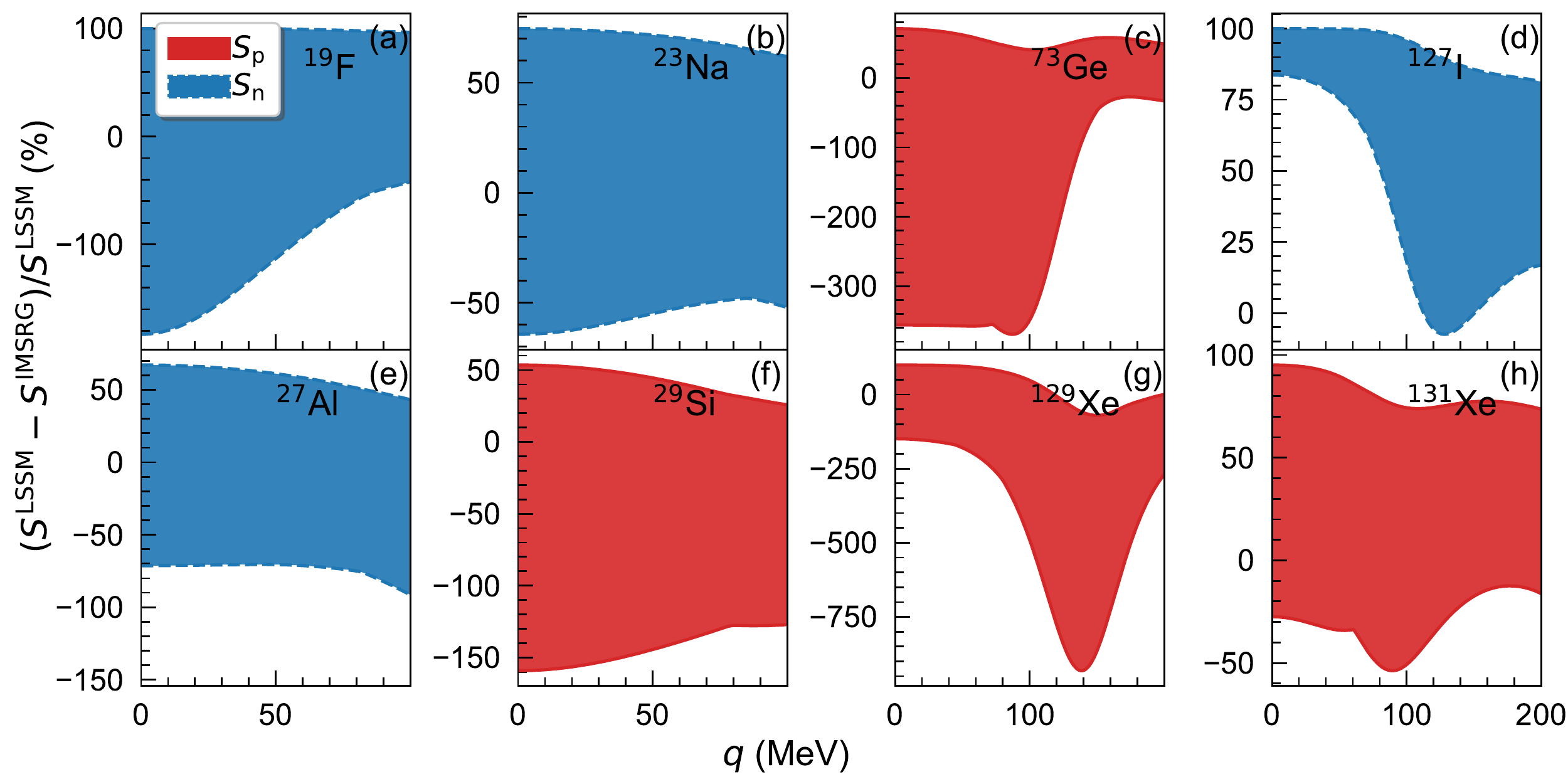}
\caption{\label{rate_non-dominant}
Discrepancy of non-dominant structure factors between LSSM and VS-IMSRG. 
All calculations are same as in Fig.~\ref{all-shell}.}
\end{figure*}

\begin{figure*}
	\begin{minipage}{0.98\columnwidth}
		\includegraphics[width=1.0\textwidth]{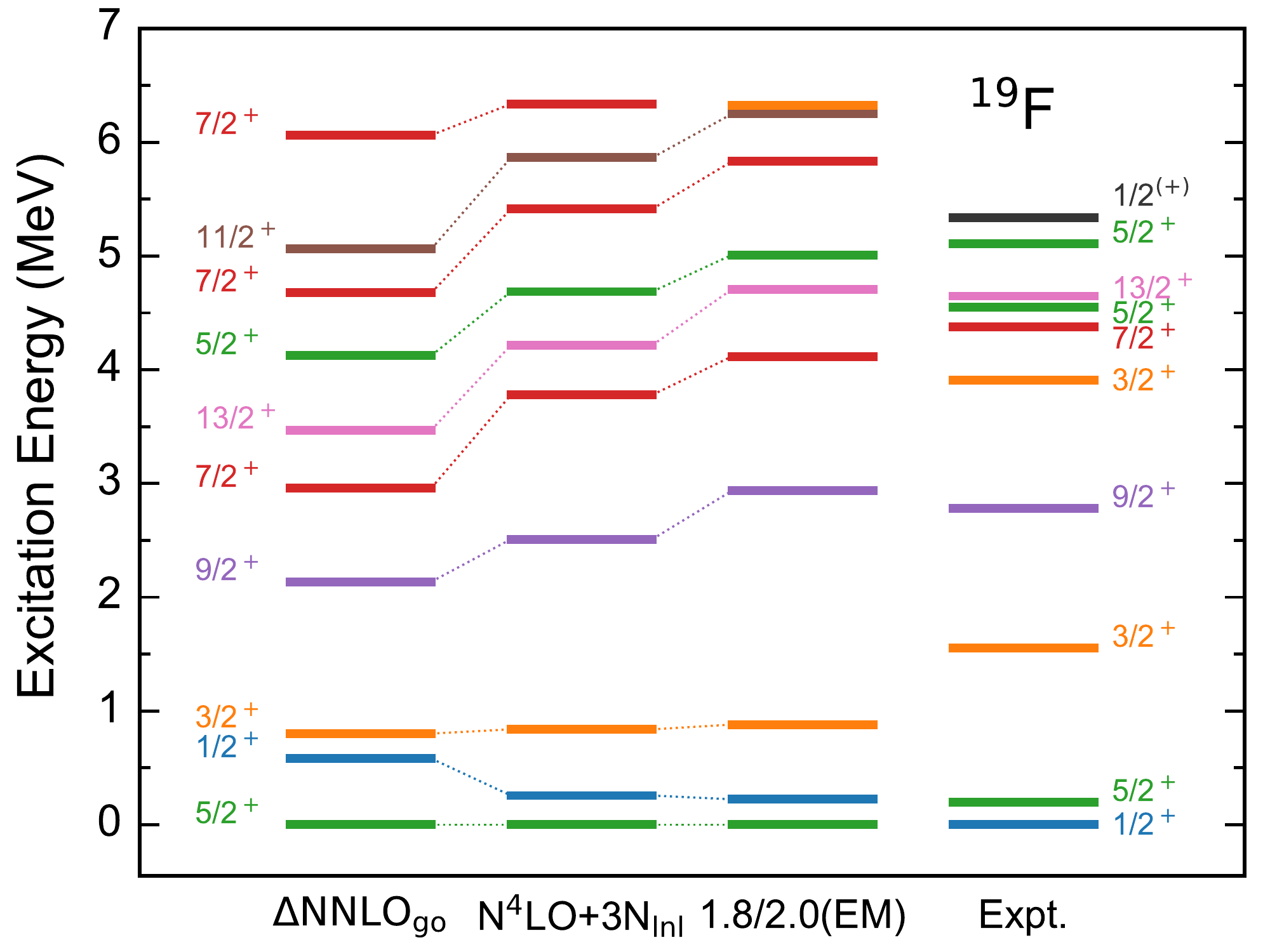}
		\caption{\label{spectra_F19} $^{19}$F spectra calculated by ab initio VS-IMSRG with different interactions, compared with experimental data \cite{nndc}.}
	\end{minipage}\hfill
	\begin{minipage}{0.98\columnwidth}
		\includegraphics[width=1.0\textwidth]{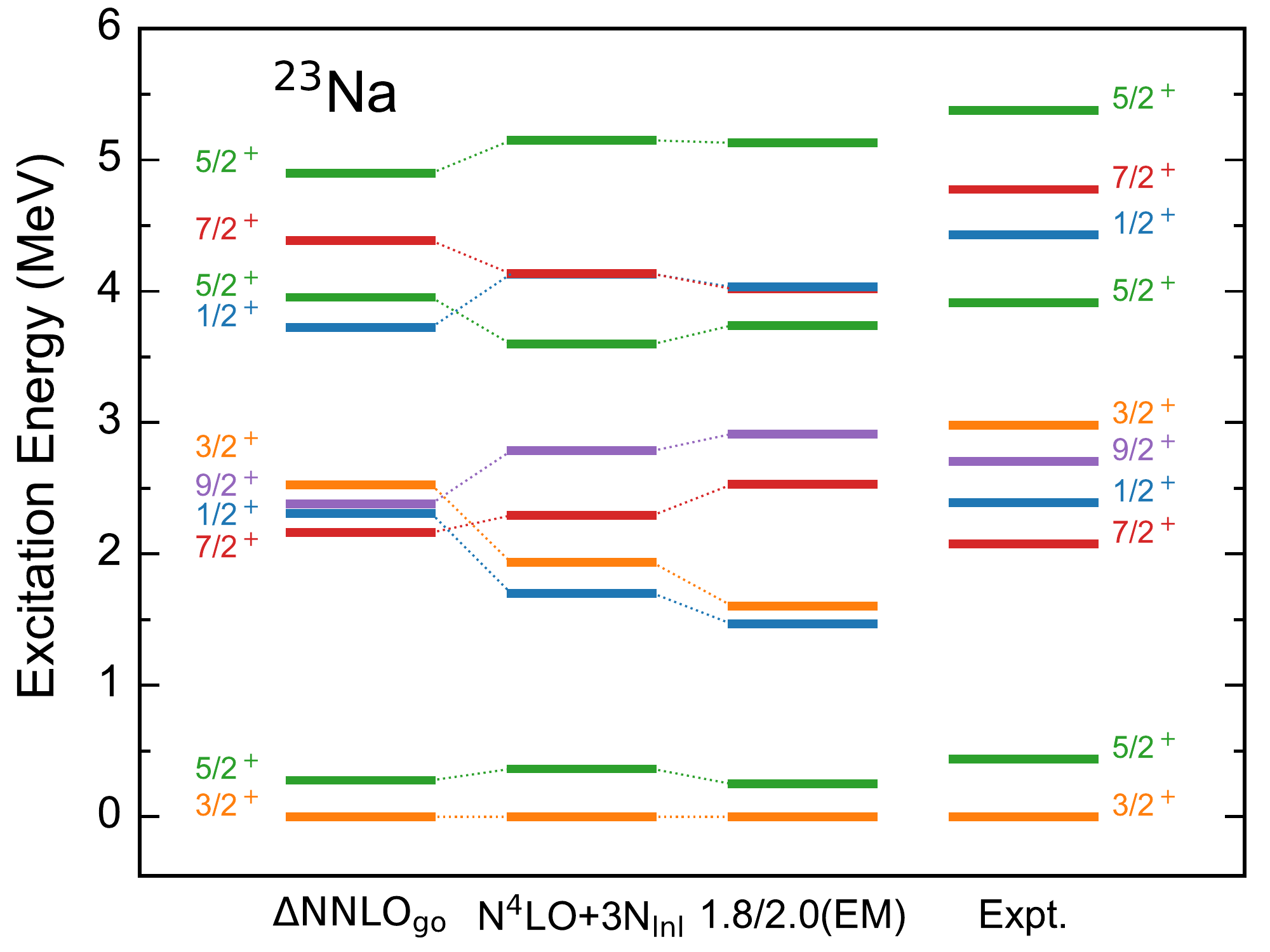}
		\caption{\label{spectra_Na23} Similar to Fig.~\ref{spectra_F19}, but for $^{23}$Na. }
	\end{minipage}
\end{figure*}

\begin{figure*}
	\begin{minipage}{0.98\columnwidth}
		\includegraphics[width=1.0\textwidth]{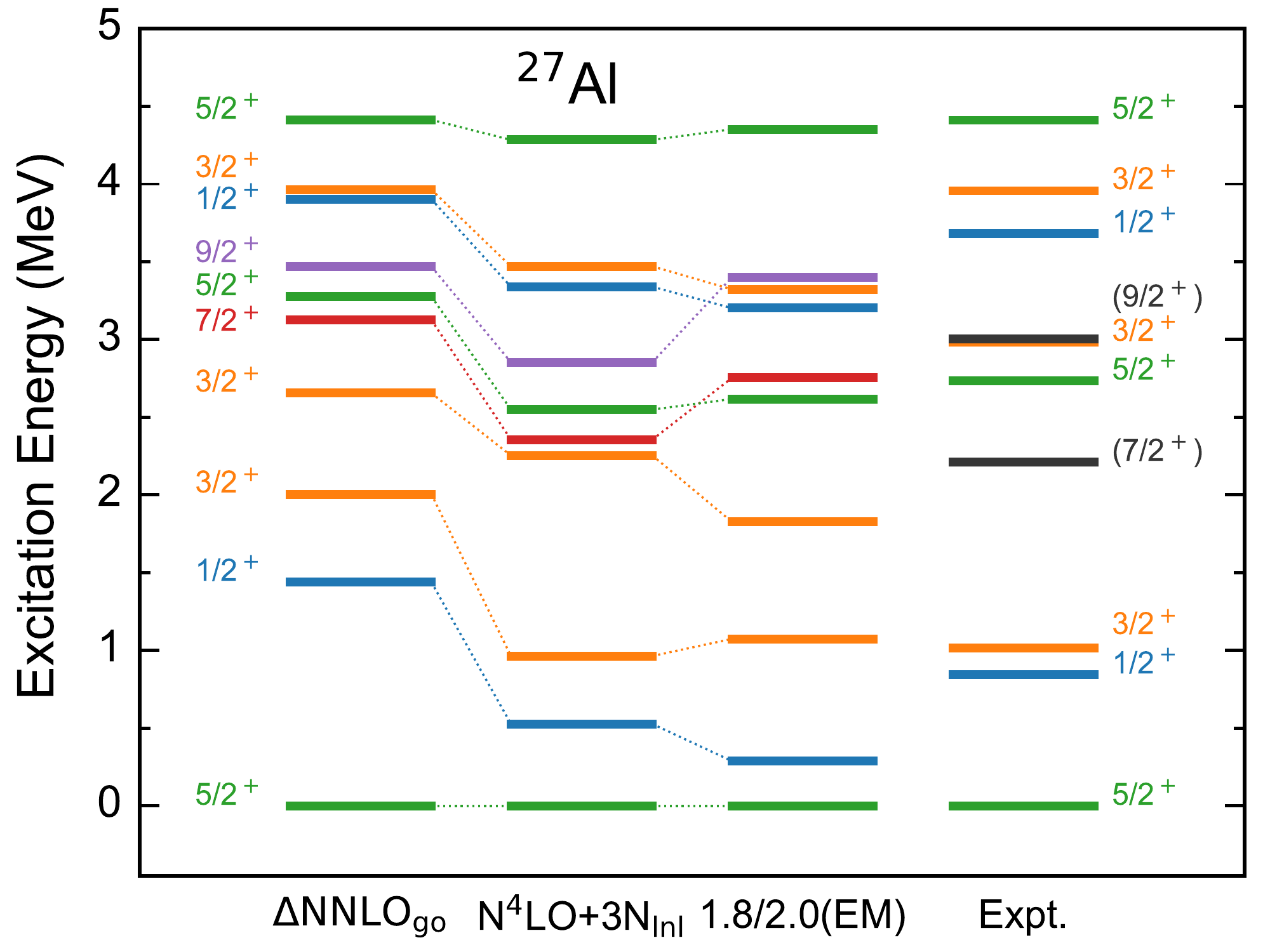}
		\caption{\label{spectra_Al27} Similar to Fig.~\ref{spectra_F19}, but for $^{27}$Al. }
	\end{minipage}\hfill
	\begin{minipage}{0.98\columnwidth}
		\includegraphics[width=1.0\textwidth]{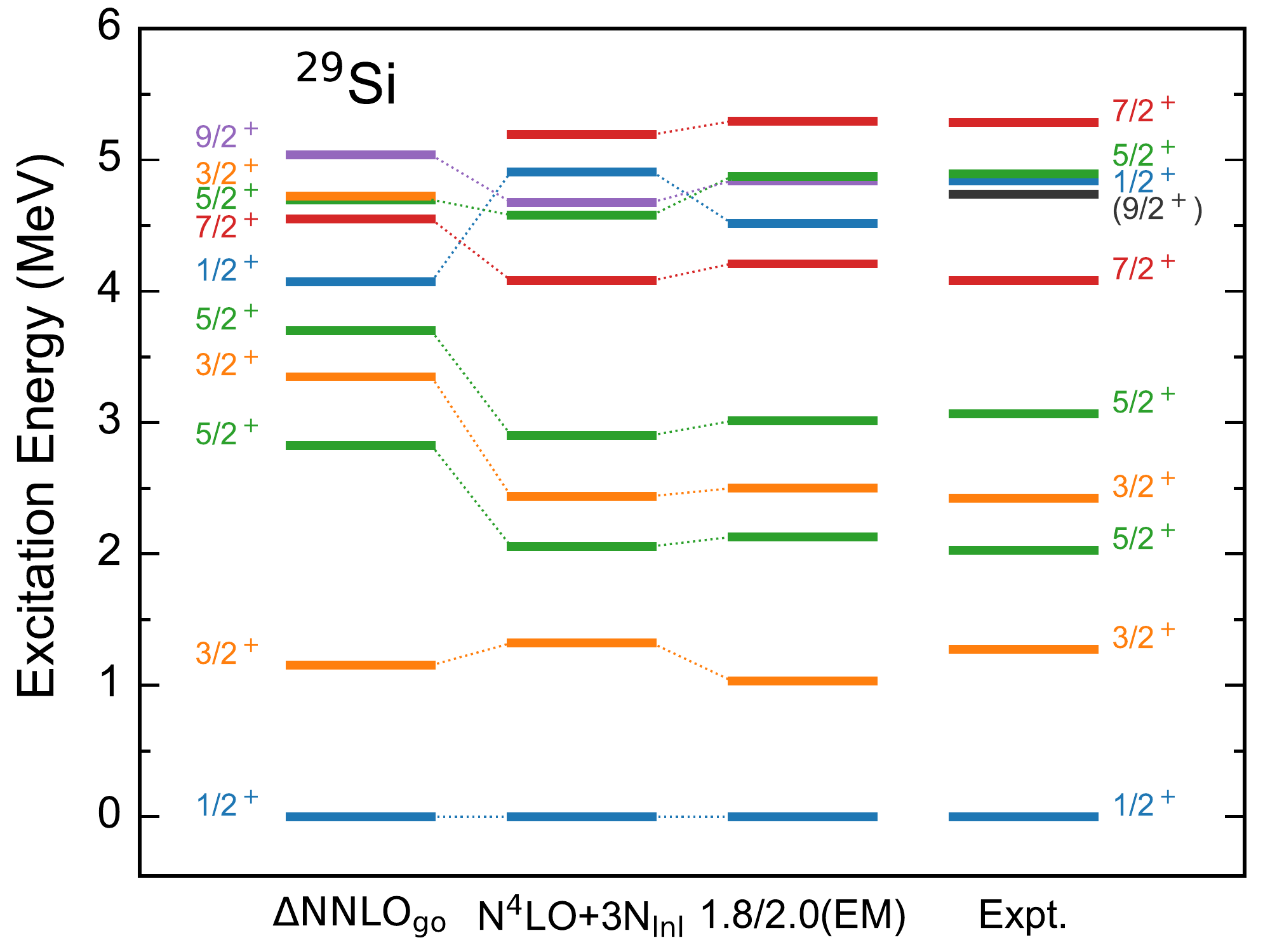}
		\caption{\label{spectra_Si29} Similar to Fig.~\ref{spectra_F19}, but for $^{29}$Si. }
	\end{minipage}
\end{figure*}

\begin{figure*}
	\begin{minipage}{0.98\columnwidth}
		\includegraphics[width=1.0\textwidth]{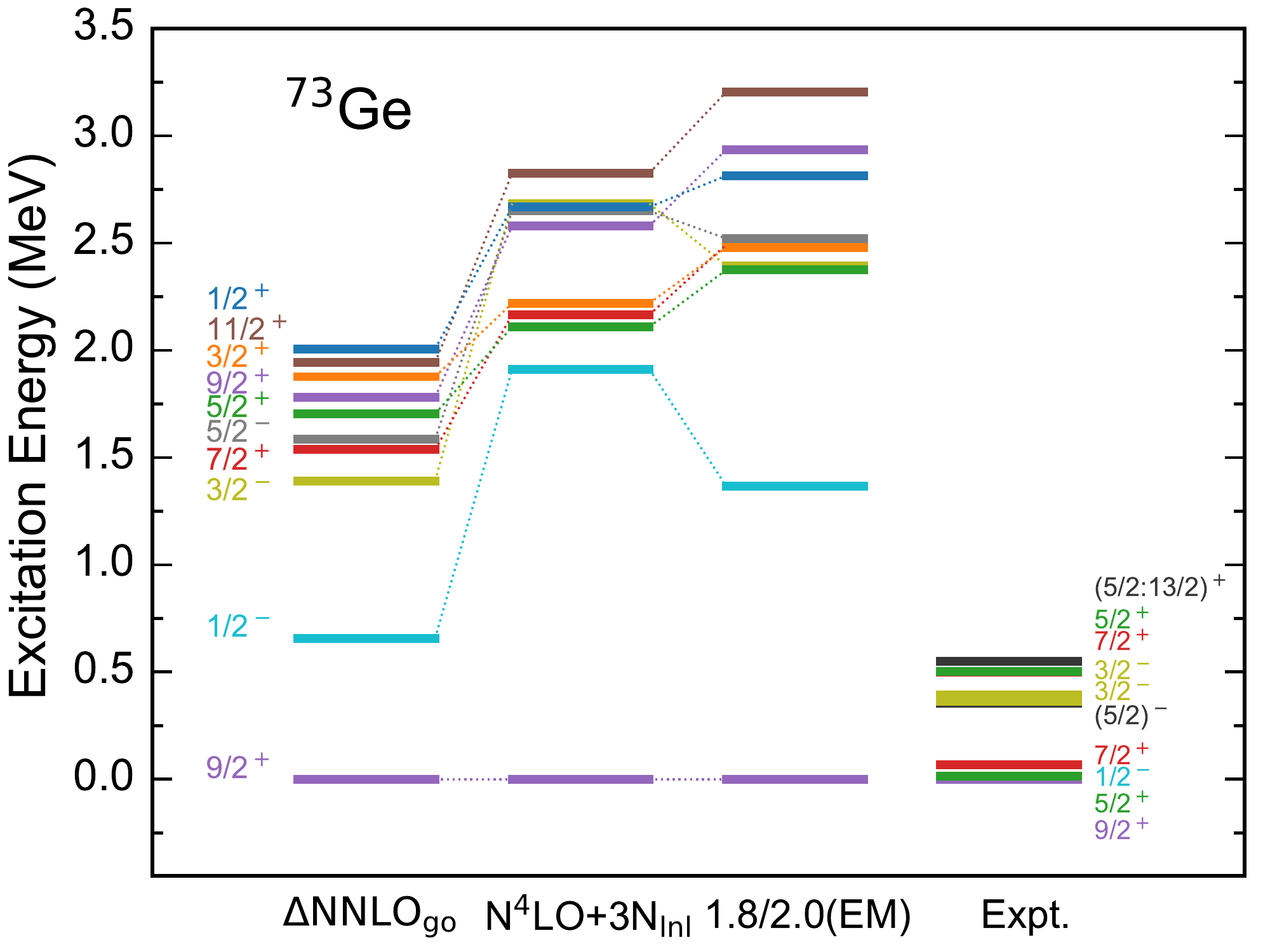}
		\caption{\label{spectra_Ge73} Similar to Fig.~\ref{spectra_F19}, but for $^{73}$Ge. }
	\end{minipage}\hfill
	\begin{minipage}{0.98\columnwidth}
		\includegraphics[width=1.0\textwidth]{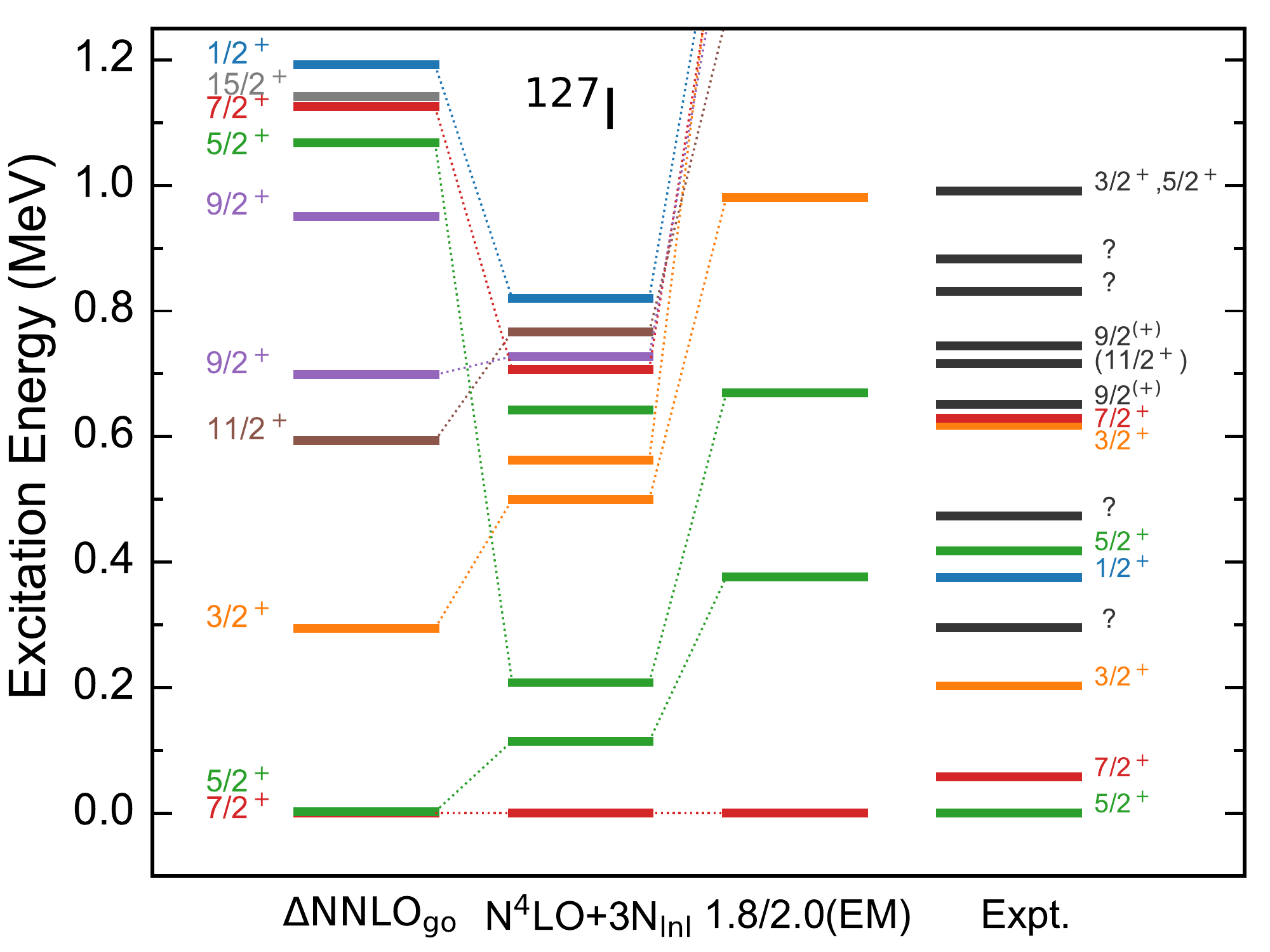}
		\caption{\label{spectra_I127} Similar to Fig.~\ref{spectra_F19}, but for $^{127}$I. }
	\end{minipage}
\end{figure*}

\begin{figure*}
	\begin{minipage}{0.98\columnwidth}
		\includegraphics[width=1.0\textwidth]{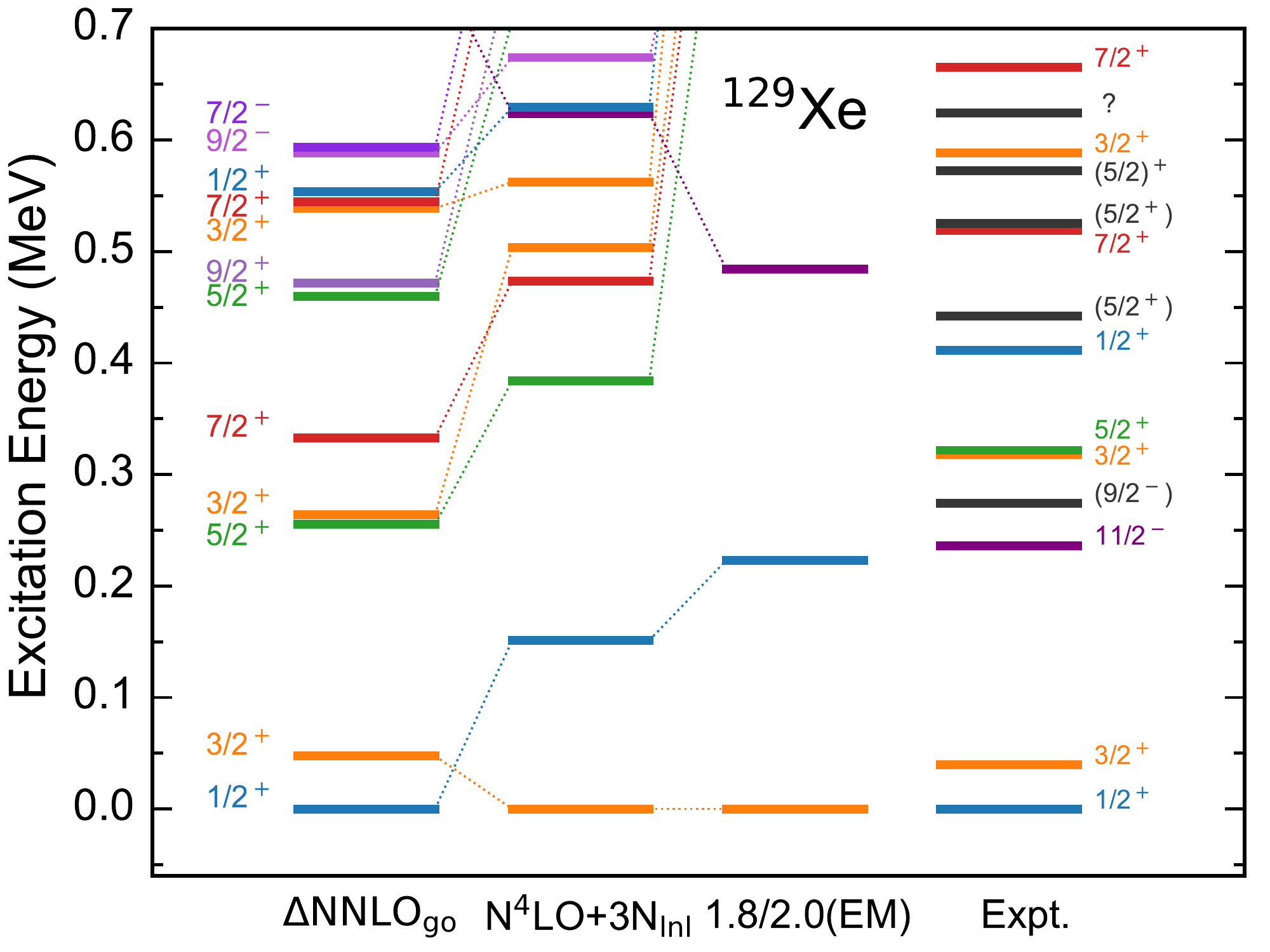}
		\caption{\label{spectra_Xe129} Similar to Fig.~\ref{spectra_F19}, but for $^{129}$Xe. }
	\end{minipage}\hfill
	\begin{minipage}{0.98\columnwidth}
		\includegraphics[width=1.0\textwidth]{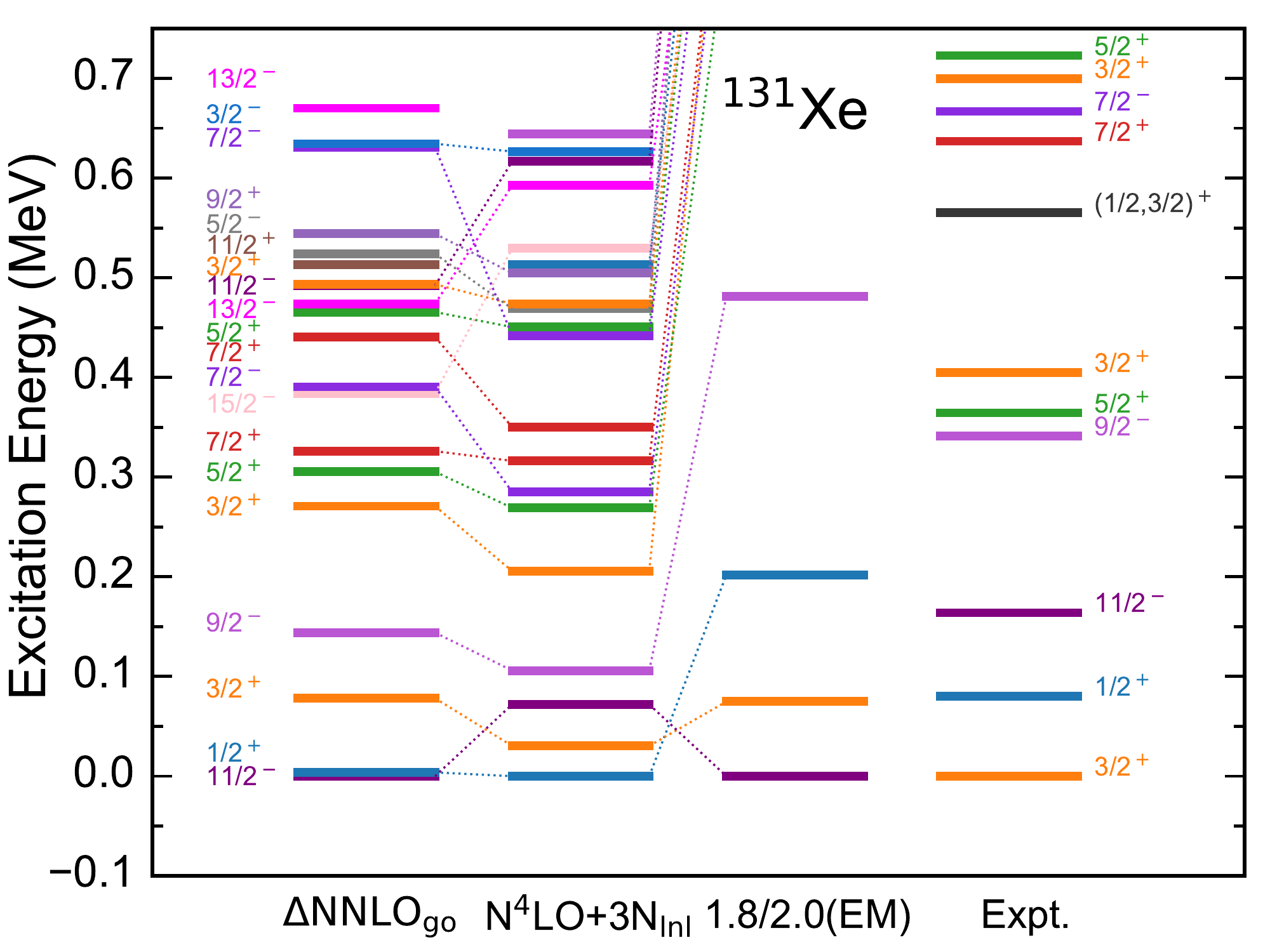}
		\caption{\label{spectra_Xe131} Similar to Fig.~\ref{spectra_F19}, but for $^{131}$Xe. }
	\end{minipage}
\end{figure*}

Figures~\ref{spectra_F19}-\ref{spectra_Xe131} show the spectra of all studied target nuclei calculated from the VS-IMSRG using the different chiral interactions discussed above. 
In order to make the diagonalization feasible for $^{127}$I and $^{129}$Xe, we allow at most eight valence particles in the neutron 0$h_{11/2}$ orbit. Note that the results of all nuclei in main text are obtained from exact diagonalization.



\end{document}